%% file: ieee_bondingcurves.tex
\documentclass[journal,onecolumn]{IEEEtran}
\IEEEoverridecommandlockouts
\usepackage{cite}
\usepackage{amsmath,amssymb,amsfonts}
\usepackage{algorithmic}
\usepackage{graphicx}
\usepackage{textcomp}
\usepackage{xcolor}
\usepackage{acronym}
\usepackage{authblk}
\usepackage{url}
\usepackage{verbatim}

\usepackage{multirow}
\usepackage{tabularx}
\usepackage{rotating}
\usepackage{makecell}
\usepackage{longtable}
\usepackage{array}
\usepackage{lscape}
\usepackage{subcaption}
\usepackage{multirow}

\newcommand{\italicparagraph}[1]{{\textnormal{\textit{#1:}}}}

\def\BibTeX{{\rm B\kern-.05em{\sc i\kern-.025em b}\kern-.08em
		T\kern-.1667em\lower.7ex\hbox{E}\kern-.125emX}}
\begin{document}

     \title{Automated Market Makers in Cryptoeconomic Systems: A Taxonomy and Archetypes}

    \author[1,2]{Daniel Kirste}
    \author[3]{Kannengießer Niclas}
    \author[2]{Lamberty Ricky}
    \author[1]{Sunyaev Ali}
    \affil[1]{Technical University of Munich, Campus Heilbronn, Germany \authorcr Email: {\tt \{daniel.kirste, sunyaev\}@tum.de}\vspace{1.5ex}}
    \affil[2]{Robert Bosch GmbH, Stuttgart, Germany \authorcr Email: {\tt \{ricky.lamberty\}@bosch.com}\vspace{1.5ex}} 
    \affil[3]{Karlsruhe Institute of Technology, Karlsruhe, Germany \authorcr Email: {\tt \{niclas.kannengiesser\}@kit.edu}\vspace{1.5ex}}

	\maketitle

    % Standard input files
	\input{01_content/01_acronyms}

\input{01_content/10_abstract}
	\input{01_content/11_keywords}

    % Productive input
    \input{01_content/20_introduction}

    \input{01_content/21_background}

    \input{01_content/22_methods}
    \input{01_content/23_taxonomy}

    \input{01_content/24_archetypes}
    \input{01_content/25_disclusion}

	\bibliographystyle{IEEEtran}
	\bibliography{01_content/00_bibliography}
 \input{01_content/26_appendix}

\end{document}

%% file: 01_content/01_acronyms.tex
\newacro{AMM}{Automated Market Maker}
\newacro{DEX}{Decentralized Exchange}
\newacro{DLT}{Distributed Ledger Technology}
\newacro{EoT}{Economy of Things}
\newacro{ICO}{Initial Coin Offering}
\newacro{IoT}{Internet of Things}
\newacro{IPO}{Initial Public Offering}
\newacro{LMSR}{Logarithmic Market Scoring Rule}
\newacro{PoS}{Proof of Stake}
\newacro{PoW}{Proof of Work}
\newacro{SEC}{United States Securities and Exchange Commission}

%% file: 01_content/10_abstract.tex
\begin{abstract}
Designing automated market makers (AMMs) is crucial for decentralized token exchanges in cryptoeconomic systems. At the intersection of software engineering and economics, AMM design is complex and, if done incorrectly, can lead to financial risks and inefficiencies.
We developed an AMM taxonomy for systematically comparing AMM designs and propose three AMM archetypes that meet key requirements for token issuance and exchange. This work bridges software engineering and economic perspectives, providing insights to help developers design AMMs tailored to diverse use cases and foster sustainable cryptoeconomic systems.

\end{abstract}

%% file: 01_content/11_keywords.tex
\begin{IEEEkeywords}
  automated market maker, decentralized finance, blockchain, cryptoeconomic system, blockchain
\end{IEEEkeywords}

%% file: 01_content/20_introduction.tex
\section{Introduction}

What if financial markets could operate without relying on central intermediaries like banks to execute trades? 
Since the development of the Bitcoin system, automated market makers (AMMs) have turned this vision into reality, reshaping how assets are traded in markets of cryptoeconomic systems---systems that leverage cryptographic principles to enable exchanges of ownership of assets based on digital tokens~\cite{zargham_curved_2020, han_how_2022}. Unlike traditional market makers, AMMs are (quasi-)autonomous software agents that provide liquidity to markets in cryptoeconomic systems and state token prices using publicly visible mathematical functions encoded in smart contracts~\cite{pourpouneh_automated_2020, drossos_automated_2025} to enable frictionless trading.
Yet, designing effective AMMs is complex.

Since the first uses in decentralized token exchanges offered by Curve Finance and Uniswap, AMMs have expanded to diverse use cases with different requirements.
For example, stablecoin exchanges (e.g., Curve Finance) require maintenance of tight price ranges and minimized price impact.
In contrast, general-purpose exchanges (e.g., Uniswap or DODO) require robust price discovery and flexible liquidity provisioning at changing price ranges.
Such requirements necessitate customized AMM designs that vary in aspects such as supported token pairs, price discovery, and liquidity provisioning.
Due to such conflicting requirements, no single AMM design can suit all use cases simultaneously.

AMM development is at the intersection of software engineering and economics, posing complex challenges in smart contract development \cite{kannengiesser_challenges_2021}, incentivization of market participants \cite{drossos_automated_2025}, and market quality~\cite{kirste_influence_2024}.
For example, AMMs must ensure that sufficient liquidity is available and stated token prices are accurate. Insufficient liquidity leads to stale prices, excessive slippage, and volatility.
Insufficient pricing mechanisms can cause financial losses due to misvalued tokens.
Beyond refining price discovery, liquidity management, and trade efficiency, AMMs must meet core requirements in software architecture \cite{xu_sok_2023}, market participant incentives \cite{drossos_automated_2025}, and overall market quality \cite{kirste_influence_2024}.
Meeting such requirements for use case--specific AMM designs requires thorough understanding of the AMM design space.

Previous works on AMMs present important design characteristics, such as liquidity sensitivity and path independence~\cite{xu_sok_2023, bartoletti_theory_2022, angeris_improved_2020, mohan_automated_2022}.
Most of these studies, however, focus on constant function market makers~\cite{xu_sok_2023, bartoletti_theory_2022, mohan_automated_2022}, neglecting designs of proactive market makers~\cite{dodoex_dodo_2023, dai_dodo_2020}, LMSR market makers~\cite{hanson_combinatorial_2003, hanson_logarithmic_2007, abernethy_optimization-based_2011}, and supply-sovereign automated market makers~\cite{zargham_curved_2020, kirste_undergirding_2025}.
As a result, it remains unclear how core design characteristics extend across AMM designs, such as those related to how AMMs source liquidity and discover adequate token prices.
The absence of a unified conceptualization complicates comparison of AMM designs, making it challenging to identify the most suitable AMM designs for use cases.
To support development of AMMs that tackle these challenges, a conceptualization of AMM designs is needed that facilitates direct comparisons between AMMs and illuminates the various ways in which AMMs ensure sufficient liquidity and adequate price discovery.

\newpage
We ask the following research questions:

\textit{\textbf{RQ1:} What are the key characteristics of AMM designs?}

\textit{\textbf{RQ2:} What are common AMM designs for use cases in cryptoeconomic systems?}

We applied a three-step research approach. First, we developed an AMM taxonomy~\cite{nickerson_method_2013} through an analysis of 122 scientific publications and 110 real-world AMMs, including Curve Finance, DODO, and Uniswap~v2.
Second, we used the AMM taxonomy to identify important design characteristics in terms of ensuring sufficient liquidity and adequate price discovery. Based on the AMM taxonomy, we developed AMM archetypes, implementing different mechanisms to source liquidity and discover adequate prices.
Third, we mapped the AMM archetypes to principal use cases.

The primary purpose of this work is to support the development of AMMs that can meet requirements for software systems and economics.
In particular, this work has the following main contributions.
First, we present an AMM taxonomy that offers a conceptual foundation for AMM development. The AMM taxonomy can guide AMM development by pointing out key dimensions and characteristics in AMM designs and offering design options to meet software requirements. This enables engineers to systematically compare AMM designs, identify trade-offs between different designs, and make informed decisions in building AMMs optimized for decentralized exchange systems.
Second, by clarifying influences of different characteristics of AMM designs, this work bridges the previously disconnected perspectives of software engineering and economics in the context of AMMs. This can help developers better meet use case requirements for AMMs.
Third, we present distinct AMM archetypes: Price-discovering LP-based AMM, Price-adopting LP-based AMM, and Price-discovering Supply-sovereign AMM. The AMM archetypes serve as software blueprints that can guide developers in designing custom AMMs suitable for use cases.

The following is organized into six sections.
Section~\ref{lbl:background} introduces cryptoeconomic systems, AMMs, and principal use cases of AMMs. Moreover, we elucidate the general software design of AMMs.
In Section~\ref{lbl:methods}, we describe how we developed the AMM taxonomy and the AMM archetypes.
In Section~\ref{sec:amm-taxonomy}, we present the AMM taxonomy and demonstrate its applicability based on 110 AMMs.
Section~\ref{sec:amm-archetypes} presents the developed AMM archetypes and their common uses. 
In Section~\ref{lbl:discussion}, we discuss our principal findings and explain the contributions and limitations of this work. Additionally, we outline future research directions on AMMs.
In Section~\ref{lbl:conclusion}, we conclude with our key takeaways.

%% file: 01_content/21_background.tex
\section{Background and Related Research}
\label{lbl:background}

To help the reader understand the aspects of AMMs and cryptoeconomic systems relevant to this work, this section describes the foundations of cryptoeconomic systems, the key software components of AMMs, and their primary purposes.

\subsection{Cryptoeconomic Systems and Distributed Ledger Technology}

Cryptoeconomic systems (e.g., based on the Bitcoin system or the Ethereum system) are sociotechnical systems that enable agents (e.g., individuals, organizations, and software artifacts) to manage ownership of assets, including claims, rights, and securities, by using principles of cryptographic systems~\cite{zargham_curved_2020, lamberty_efficiency_2023}. In this section, we introduce the foundations of cryptographic and economic systems combined in cryptoeconomic systems. Building on those foundations, we introduce DLT and its role in the operation of cryptoeconomic systems~\cite{zhang_security_2019}.

\paragraph{Cryptoeconomic Systems}
Cryptoeconomic systems allow for more decentralized operation of financial systems compared to contemporary ones commonly provided by single actors like banks~\cite{sunyaev_token_2021}. Cryptoeconomic systems combine principles of cryptographic systems and economic systems. Cryptographic systems are suites of cryptographic techniques used to reach target security levels, such as in terms of confidentiality~\cite{bell_explaining_2003, cortier_survey_2011}. The basic functions to be offered by cryptographic techniques in cryptographic systems are key generation, encryption, and decryption. Key generation algorithms produce secrets, also called keys, that can be used to encrypt and decrypt data. 
Asymmetric key techniques are commonly used to authenticate user identities in computer systems (e.g., blockchain systems) by digital signatures~\cite{goldwasser_bellare_2008, diffie_hellman_1976, Chaum_1983}.
Using digital signatures, cryptoeconomic systems authorize asset transfers. For example, market participants in stock markets must authenticate against brokers to initiate asset transfers.

Economic systems are social systems in which market participants exchange resources in the form of products and services.
A prevalent form of economic system in modern times is the market economy~\cite{hurwicz_design_1973}. In market economies, prices and production are determined by the interaction of supply and demand from all market participants (e.g., producers, consumers, investors, and traders)~\cite{samuelson2009}. Market participants exchange assets, such as goods and services.
To enable asynchronous asset exchanges, exchanges operate order books that record buy and sell offers and settle trades when a matching counterparty is found. There are two types of orders. First, limit orders refer to instructions to buy or sell assets at a specified price but without the guarantee of immediate execution~\cite{abergel_limit_2016}. Limit orders are stored in order books. As the counterpart to limit orders, market orders are instructions to buy or sell assets immediately at a given price~\cite{abergel_limit_2016}.
Immediately available volumes to settle market orders (e.g., through limit orders) corresponds to the liquidity available in a market~\cite{hirshleifer_liquidity_1971}.

Market makers enable smooth asset trading by providing liquidity to markets. A market maker is a rational market participant who quotes bid (buy) and ask (sell) prices for trading pairs~\cite{ohara_microeconomics_1986}. A trading pair refers to two types of assets that can be traded against each other, for example, Bitcoin against USD or USD against Wheat.
Market makers place limit orders in the order book committing their willingness to trade at bid/ask prices to market participants, consequently providing liquidity to the market~\cite{ohara_microeconomics_1986}.

Market makers determine bid/ask prices using various methods, including pricing functions that aggregate market quotes. Such a pricing function can compute the market maker's asset prices by averaging all bid/ask quotes of all investors. The pricing function of market makers is usually private and not known to other market participants~\cite{ohara_microeconomics_1986, hendershott_market_2007}.
Market makers leverage bid/ask spreads by buying and selling assets with added surcharges~\cite{pourpouneh_automated_2020, hendershott_market_2007}. 

As a prerequisite for asset exchanges, market makers must hold balanced amounts of all assets in their inventory that are offered in trading pairs to market participants~\cite{ohara_microeconomics_1986}. Market makers ideally sell a number of assets (e.g., USD) and simultaneously buy the equivalent number of assets of the same kind. When arbitrageurs and investors buy more underpriced assets of one kind, the market maker sells more of this asset than it buys.
This makes market makers subject to inventory imbalances that can render market makers unable to trade asset pairs~\cite{hendershott_price_2014, hendershott_market_2007}.

\paragraph{Cryptoeconomic Systems based on Distributed Ledger Technology}

Distributed ledger technology (DLT), including blockchain technology \cite{kannengieser_trade-offs_2020}, is often used to operate the infrastructure of cryptoeconomic systems~\cite{zargham_curved_2020, sunyaev_token_2021}; in particular, to record token transfers and authenticate market participants using cryptographic techniques.
DLT enables the operation of distributed ledgers, a kind of distributed database that processes and records (financial) transactions based on logic manifested in software code.

DLT systems are governed by stakeholders who decide the rules for creation, allocation, and distribution of digital tokens that represent assets. 
Tokens are typically specified and managed in smart contracts (e.g., ERC-20 standard) that map token balances to unique identifiers of market participants (e.g., addresses of externally owned accounts in the Ethereum system)~\cite{ethereum_foundation_erc-20_2023}. Smart contracts are software programs that allow for the automated execution of transaction logic~\cite{Szabo_2005, kannengiesser_challenges_2021}. Transactions can manipulate the token mapping, enabling transfers of asset ownership.

\subsection{Automated Market Makers}
\label{sec:amm-foundation}

AMMs are market makers implemented as software agents that trade tokens with market participants at self-determined prices in an automated manner. In contrast to conventional market makers (e.g.,~trading organizations), AMMs settle tokens based on trading strategies that are defined in mathematical formulas encoded in smart contracts.
Because smart contract code is visible to users of DLT systems \cite{kannengiesser_challenges_2021}, trading strategies of AMMs is visible to market participants with read access to the ledger \cite{drossos_automated_2025}.

\begin{figure}[h]
    \centering
    \includegraphics[width=0.65\textwidth]{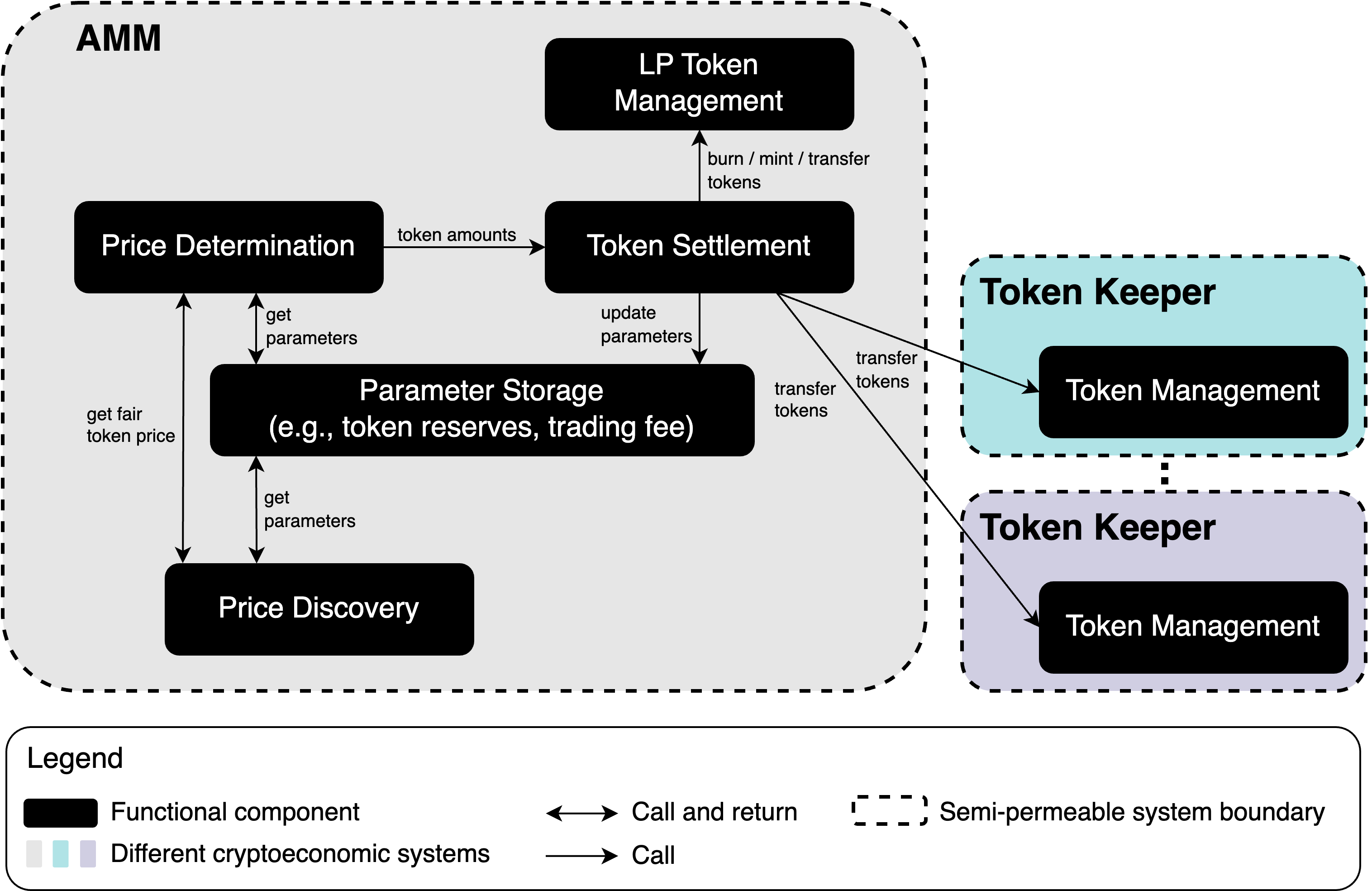}    
    \caption{Illustration of AMM component of Uniswap~v2~\cite{zinsmeister_uniswap_2020}}
    \label{fig:amm-components}
\end{figure}

AMM designs commonly include a \textit{price discovery component}, \textit{price determination component}, a \textit{parameter component}, a \textit{token settlement component}, at least one \textit{token management component}, and a \textit{liquidity provider (LP) token management component}. Figure \ref{fig:amm-components} illustrates the design of the AMM used in Uniswap~v2.

The \textit{price discovery component} implements the logic to discover token prices. Typically, the price discovery component is part of the AMM and uses parameters of the parameter component for price discovery. The preliminary token price is passed to the price determination component.

The \textit{price determination component} implements a price determination mechanism to compute the bid/ask token prices. The price determination component adjusts token prices based on the parameters of the parameter component. Market participants can trade tokens at the bid/ask prices determined by the price determination component.
When market participants initiate transactions, the price determination component calculates the amount of tokens the market participant buys in return for the amount of tokens sold to the AMM. Both token amounts are passed to the token settlement component.

The \textit{parameter component} stores parameters the AMM uses to determine prices, settle transactions, and govern parameter changes of the AMM. Exemplary parameters of the parameter component are trading fees, amount of token reserves, token weights, and recent prices \cite{zinsmeister_uniswap_2020}.
A set of parameters defines the state of an AMM. Trades of market participants with the AMM trigger state transitions.
The price discovery and price determination components calculate the amount of tokens the market participant will receive in return for the provided amount of tokens. The calculation is based on the parameters in the AMM, the input parameters of the transaction, and parameters external from the AMM. The token price results from the amount of received tokens divided by the amount of provided tokens (from the market participant's perspective).

The \textit{token settlement component} calls the token management component of token keepers to initiate the actual token transfer through a token keeper.
A token keeper is a software agent, often implemented as a smart contract, that controls the token management component. Token keepers are part of at least one cryptoeconomic system. They are often external to AMMs. AMMs can be token keepers themselves.
Within each token keeper, a \textit{token management component} manages the tokens of market participants, including AMMs in inventories. Token management components maintain account books that map token balances to unique identifiers (e.g., account addresses) of market participants. \textit{Transfer}, \textit{mint}, and \textit{burn} transactions update the account book in the token management component. 

AMMs have token inventories (liquidity pools) managed by at least one token keeper
~\cite{zinsmeister_uniswap_2020, xu_sok_2023, angeris_improved_2020}. To settle transactions, AMMs instruct token keepers that manage tokens involved in the transactions to transfer tokens by updating their account books. For example, a market participant exchanges 1~WETH for 1000~USDC. To settle this transaction, the AMM instructs the ERC-20 smart contract of WETH to transfer 1~WETH from the market participant to the AMM's liquidity pool. In the ERC-20 smart contract of WETH, the market participant's token balance is subtracted by one.
The token balance of the AMM's liquidity pool is increased by one. Vice versa, the ERC-20 smart contract of USDC is instructed to increase the market participant's token balance by 1,000 and decrease the token balance of the AMM's liquidity pool by 1,000.

The \textit{LP token management component} is an optional AMM-internal component that is used to manage tokens deposited by LPs. Market participants can deposit tokens into the liquidity pools of AMMs to become LPs~\cite{zinsmeister_uniswap_2020, aoyagi_liquidity_2020}. When depositing tokens into the liquidity pools, LPs receive LP tokens. LP tokens represent a claim on a share of the liquidity pool that allows LPs to withdraw their share. LP tokens cannot be traded via the AMM~\cite{xu_sok_2023, evans_liquidity_2021, zinsmeister_uniswap_2020}.
The LP token management component stores and administrates the account book that maps LP tokens of the AMM to the unique identifiers of the LPs.\\

While these core components form the foundation of AMM design, they can be customized to meet different requirements.
Constant function market makers implement the predominant AMM design used by Uniswap~v2~\cite{zinsmeister_uniswap_2020}, PancakeSwap~\cite{pancakeswap_docs_pancakeswap_2023}, and SushiSwap~\cite{sushi_docs_sushi_2023}, as illustrated in Figure~\ref{fig:amm-components}.
Constant function market makers use mathematical conservation functions for price discovery. The price discovery mechanism adjusts token prices based on buy and sell transactions of market participants that change the AMM's token reserves~\cite{bartoletti_theory_2022, schlegel_axioms_2022, engel_composing_2021, wang_automated_2020, pourpouneh_automated_2020, aoyagi_liquidity_2020}.
Proactive market makers, such as DODO~\cite{dodoex_dodo_2023} and WooFi~\cite{woofi_woofi_2023}, use external price oracles that incorporate price discovery components. Proactive market makers do not discover token prices on their own. Instead, they adopt token prices from external price oracles that are often operated by third parties~\cite{dodoex_dodo_2023, xu_sok_2023, bichuch_axioms_2022, dai_dodo_2020}.
The different AMM designs are necessary to suit different use cases, such as decentralized token exchanges and token issuance.

\subsection{Principal Purposes of Automated Market Makers}\label{sec:purposes}

Primary use cases of AMMs are \textit{decentralized exchange (DEX)} and \textit{token issuance}. These use cases can be further divided into six specific subcategories, as described below.

\paragraph{Decentralized Token Exchange} 
Decentralized token exchanges allow market participants to exchange tokens without the need for central authorities~\cite{meyer_decentralized_2021, schueffel_defi_2021}, such as brokers.
Decentralized token exchanges can be used to trade four token types: \textit{correlated tokens}, \textit{uncorrelated tokens}, \textit{non-fungible tokens (NFTs)}, and \textit{prediction tokens}.

\italicparagraph{Correlated Tokens} 
Correlated tokens are tokens with linked prices. When the price of one token increases (or decreases), prices of correlated tokens also increase (or decrease). Strongly correlated token pairs are supposed to be exchanged at a constant rate~\cite{egorov_stableswap_2019}. To enable cost-efficient exchanges at constant rate, the market must be highly liquid at this exchange rate~\cite{egorov_automatic_2021}. Example correlated token pairs are Circle's USDC and Tether's USDT. Both tokens are paired with each other and exchangeable at a one-to-one ratio.

\italicparagraph{Uncorrelated Tokens} 
Uncorrelated token exchanges enable trades of tokens whose token prices are mutually independent. Uncorrelated token pairs require liquidity in wide price ranges because tokens cannot be exchanged at a constant rate~\cite{mohan_automated_2022}. Instead, the exchange rate varies due to volatility and price fluctuations. An example of an uncorrelated token pair is Bitcoin paired with a stable token, such as Circle's USDC.

\italicparagraph{Non-fungible Tokens} 
Each NFT is unique and thus has an individual value. Therefore, NFTs are inherently non-interchangeable.
To enable interchangeability, all NFTs of one collection are treated equally and assumed to be interchangeable~\cite{rarity_what_2022}.
For example, NFTs of the Bored Ape Yacht Club collection are be paired with Ether. NFTs in the Bored Ape Yacht Club collection are treated equally and do not have a unique value assigned.

\italicparagraph{Prediction Tokens} 
Prediction tokens are used in prediction markets. Prediction markets enable market participants to bet on outcomes of future events, for example,  outcomes of elections or company stock prices at a specific future point in time~\cite{fountain_what_2011}. Market participants can place their bets into AMMs.
To place bets, market makers deposit tokens into an AMM's liquidity pool. The occurrence of events on which market participants have bet triggers the closure of the prediction market. When the market closes, AMMs compare event outcomes with the bets placed by market participants.
The AMM initiates payouts of tokens deposited by market participants depending on the event outcome~\cite{wang_prediction_2023, peterson_augur_2018, carvalho_permissioned_2020}. For example, market participants who bet on the outcomes of events receive tokens; other market participants lose their wagers.
Examples of prediction markets are offered by Augur~\cite{peterson_augur_2018} and Zeitgeist~\cite{zeitgeist_docs_prediction_2022}. Those prediction markets enabled market participants to bet on events such as the outcome of the 2020 U.S. presidential election.

\paragraph{Token Issuance}
Token issuance refers to the process of minting and distributing tokens to market participants. In token issuance, AMMs map the token price to the token supply~\cite{hertzog_bancor_2018}. There are two types of token issuance: \textit{curation token} and \textit{initial token offerings}.

\italicparagraph{Curation Tokens} 
Curation tokens are used to curate market participants' perceptions of the value of an asset, such as artworks, data sets, and machine learning (ML) models. AMMs are used to adjust token prices to update token prices according to market participants' perceptions continuously~\cite{rouviere_introducing_2019, the_graph_graph_2023}. For example, the Ocean Protocol offers an AMM for curation markets that issues tokens for data sets used to value the provided data sets regarding their quality of training ML models. It curates the perceptions of market participants regarding the quality of the data set~\cite{ocean_protocol_foundation_ocean_2020}.

\italicparagraph{Initial Token Offerings} 
In initial token offerings, token issuers (e.g.,~individuals and organizations) collect funding to finance endeavors by selling shares of the endeavor represented in the form of tokens~\cite{fenu_ico_2018}. Such endeavors include building new infrastructure, new Dapps, or any other projects (e.g., The Dao, Fetch.ai, Bancor)~\cite{zhao_dao_2017,simpson_fetch_2019,hertzog_bancor_2018}.
AMMs can establish thick markets that enable market participants to buy and sell tokens in the context of initial token offerings.
Initial token offerings helped finance the development of the Ethereum system in 2014~\cite{icodrops_ethereum_2018} and the Tezos system in 2017~\cite{icodrops_tezos_2017}.

%% file: 01_content/22_methods.tex
\section{Methods}
\label{lbl:methods}
We applied a three-step approach to uncover the characteristics of AMM designs and develop AMM archetypes.
First, we developed an AMM taxonomy~\cite{nickerson_method_2013} based on literature and AMM implementations. 
Second, we used the AMM taxonomy to develop AMM archetypes. 
Third, we analyzed scientific publications, gray literature, and AMM implementations to map the AMM archetypes to use cases.

\subsection{AMM Taxonomy Development}
\label{sec:amm-taxonomy-method}

We developed an AMM taxonomy~\cite{nickerson_method_2013} to uncover differences between AMM designs.
First, we defined the meta-characteristic \textit{AMM design characteristics} to scope the taxonomy development.
Second, we specified ending conditions (see Table~\ref{tab:methods-ending-cond}) that helped us recognize when the taxonomy reached sufficient quality to terminate the development.
Third, we applied the conceptual-to-empirical and empirical-to-conceptual approaches (see Table~\ref{tab:methods-tax-overview}) in five iterations to develop the taxonomy.
While we analyzed scientific and non-scientific publications on AMMs in the conceptual-to-empirical approach, we analyzed real-world implementations AMMs to extract characteristics of AMM designs.
In total, we analyzed 122 publications in the conceptual-to-empirical approach and 110 implementations of AMMs in the empirical-to-conceptual approach.
The following section details each iteration in the taxonomy development process.

\begin{table}[b]
    \centering
    \caption{Ending conditions for the taxonomy development}
    \label{tab:methods-ending-cond}
\begin{tabularx}{\textwidth}{|c|l|X|}
\hline
\textbf{Type} & \textbf{Name} & \textbf{Description} \\ 
\hline
\multirow{5}[10]{*}{Objective}
& Exhaustiveness & The characteristics and dimensions exhaustively describe AMM designs\\
\cline{2-3}
& Mutual Exclusiveness & The characteristics and dimensions do not overlap in semantics\\
\cline{2-3}
& Relevance & Each characteristic of each dimension is required for the classification of at least one AMM design in the taxonomy\\
\cline{2-3} 
& Representativeness & A sufficient selection of publications and AMMs representative of AMM designs were incorporated into the taxonomy \\
\cline{2-3} 
& Robustness & No changes were made to the taxonomy in the last iteration\\
\hline
Subjective & Conciseness & The taxonomy includes a limited number of relevant dimensions and characteristics to describe AMM designs \\
\hline
\end{tabularx}
\end{table}

\begin{table*}[h]
    \renewcommand{\arraystretch}{1.3}
    \caption{Overview of the development of the AMM taxonomy}
    \label{tab:methods-tax-overview}
    \centering
        \begin{tabularx}{\textwidth}{|p{2cm}|X|c|c|c|c|c|c|}
        \hline
            \textbf{Approach} & \textbf{Type} & \textbf{Iter. 0} & \textbf{Iter. 1} & \textbf{Iter. 2} & \textbf{Iter. 3} & \textbf{Iter. 4} & \textbf{Summary}\\
            \hline
            \multirow{3}{*}{\shortstack[l]{Conceptual-\\to-empirical}} 
                & Confirmatory publications
                & 5
                & 25
                & n.a.
                & 63
                & n.a.
                & 93 \\
\cline{2-8}
                & Conflicting publications
                & 14
                & 8
                & n.a.
                & 7
                & n.a.
                & 29\\
\cline{2-8}
                & Overall publications 
                & 19
                & 33
                & n.a.
                & 70
                & n.a.
                & 122 \\
            \hline
            \multirow{3}{*}{\shortstack[l]{Empirical-to-\\conceptual}} 
                & Confirmatory AMMs
                & n.a.
                & n.a.
                & 46
                & n.a.
                & 47
                & 93\\
\cline{2-8}
                & Conflicting AMMs
                & n.a.
                & n.a.
                & 14
                & n.a.
                & 3
                & 17 \\
\cline{2-8} 
                & Overall AMMs
                & n.a.
                & n.a.
                & 60
                & n.a.
                & 50
                & 110\\
        
            \hline
            \multicolumn{8}{r}{
\textit{n.a.: not applicable}} \quad \textit{Iter.: Iteration}\\
        \end{tabularx}
\end{table*}

\paragraph{Conceptual-to-Empirical (\textit{Iteration 0})} To develop an initial version of the AMM taxonomy from literature, we use the conceptual-to-empirical approach.
To assess the relevance of publications on AMMs, we defined inclusion criteria (see Table~\ref{tab:methods-inclusion}). We considered a publication relevant only if it met all inclusion criteria.
Subsequently, we started to compile a set of potentially relevant publications. We first selected publications we deemed particularly relevant for developing the AMM taxonomy. This selection resulted in an initial set of 31 potentially relevant publications, including peer-reviewed and gray literature.
After applying the inclusion criteria, we excluded twelve publications that did not present sufficient details of AMM designs. 
The final set of relevant literature comprised 19 publications.

After the literature search, we read the full texts of the 19 publications. Then, we analyzed them to extract dimensions and characteristics describing AMM designs through coding.
In the coding, we recorded a name, description, original source, and preliminary characteristics of AMM designs.
The initial coding resulted in 41 preliminary characteristics that we associated with 31 preliminary dimensions. We resolved ambiguities and inconsistencies between the preliminary characteristics and dimensions in three refinement rounds with intense discussions among the authors. For example, we merged the characteristics \textit{sufficient funds}, \textit{path deficiency}, and \textit{non-depletion} into \textit{path deficiency}. 
After refining the preliminary characteristics and associated dimensions, the initial version of the AMM taxonomy comprised 16 dimensions with 43 subordinate characteristics.

\begin{table}[t]
    \centering
    \caption{Inclusion criteria used in the literature search to assess the relevance of publications}
    \label{tab:methods-inclusion}
\begin{tabularx}{\textwidth}{|X|p{10.9cm}|}
\hline
\textbf{Name}      & \textbf{Description}                                 \\ 
\hline                
AMM Design Description & The publication describes at least one AMM design in sufficient detail to extract design characteristics and dimensions \\
\hline    
English Language       & The publication must be in English language    \\ 
\hline    
Topic Fit & The publication must describe at least one AMM design \\
\hline    
Uniqueness & The publication must not be already included in the set of literature\\ 
\hline                             
\end{tabularx}
\end{table}

\paragraph{Conceptual-to-Empirical (\textit{Iteration 1})}
To enrich the initial version of the AMM taxonomy, we conducted a backward- and forward search based on the previously analyzed 19 publications. The first round of backward and forward searches yielded 1,086 additional publications (i.e.,~243 publications by backward search and 843 by forward search). We applied the inclusion criteria (see Table~\ref{tab:methods-inclusion}) to the meta-information (e.g.,~title, keywords, abstract) of the potentially relevant publications. We excluded 291 duplicates and 692 publications because they were off-topic or lacked detailed descriptions of AMM designs.
We added 103 relevant publications to the set of relevant literature on AMMs for taxonomy development.

Next, we randomly selected publications to extract characteristics and dimensions through coding.
After coding each publication, we resolved ambiguities and inconsistencies in four refinement rounds. For example, we merged the dimensions \textit{liquidity reversibility} and \textit{liquidity invariance} into \textit{liquidity changeability}. 
We ended the analysis when no refinements of the AMM taxonomy were required for the last ten analyzed publications. After analyzing 52 relevant publications, the refined AMM taxonomy included 29 AMM dimensions and 68 characteristics.

\paragraph{Empirical-to-Conceptual (\textit{Iteration 2})}
To test the AMM taxonomy, we analyzed real-world AMM implementations.
We selected a sample of 100 AMMs with the largest 24-hour trading volume reported on \textit{www.coinmarketcap.com}. When AMMs had identical designs but deployments to different DLT systems (e.g., Uniswap v3 in the Ethereum system, Uniswap v3 in the Polygon system), we selected only the AMM with the largest trading volume for analysis. We treated different versions of AMMs as separate designs (e.g.,~Uniswap~v2 and Uniswap~v3). We excluded protocols that do not implement AMMs, such as order-book, derivative, and aggregator protocols.
To enrich the data set, we added ten AMMs whose designs strongly differ from those of the 100 AMMs selected from \textit{www.coinmarketcap.com} to increase the exhaustiveness of the AMM taxonomy.

To classify the selected AMMs in the AMM taxonomy, we used official documentation, whitepapers, yellowpapers, and code repositories. If such official sources lacked necessary information, we extended the search to gray literature and contacted the developers to gather the necessary information. We selected the 60 AMMs with the largest trading volume from the set of 110 AMMs for AMM classification and used the remaining 50 AMMs to test the robustness of the AMM taxonomy in iteration 5.

Next, we classified the selected 60 AMMs into the AMM taxonomy. We dropped pure mathematical descriptions of (parts of) AMMs because they did not match the meta-characteristic.
For example, we removed the \textit{curve monotony}, \textit{curve scaling}, and \textit{curve differentiability} dimensions because we decided that the detailed mathematical descriptions are not concise and implicitly covered by other dimensions such as \textit{translation invariance}. 

After the classification of 60 AMMs into the taxonomy, no refinements of the AMM taxonomy were required for the last 10 analyzed AMMs. After iteration~2, the AMM taxonomy included 18 dimensions and 45 characteristics. 

\paragraph{Conceptual-to-empirical (\textit{Iteration~3})}
We applied the conceptual-to-empirical approach and extracted characteristics and dimensions from the remaining 70 publications from iteration~1.
We refined the AMM taxonomy when we recognized the need to add dimensions and characteristics to the AMM taxonomy or to redefine existing ones. For example, we merged the dimensions \textit{dynamic trading fees} into \textit{asset risk management}~\cite{milionis_automated_2022}.
The analysis of the last 34 of the 70 publications did not lead to refinements of the taxonomy. Thus, 63 of the 70 publications confirmed the preliminary AMM taxonomy.

\paragraph{Empirical-to-conceptual (\textit{Iteration 4})}
To test the robustness of the AMM taxonomy, we analyzed the 50 AMMs remaining from iteration 2 and classified them into the AMM taxonomy. For the analysis and classification, we proceeded as described in Iteration~2.
Among the 50 AMMs, three AMMs required minor refinements of the AMM taxonomy. For example, we added the AMM dimension \textit{token price source} and added \textit{constant-power-sum} to the \textit{price discovery mechanism} dimension. In this iteration, we had 47 confirmatory AMM implementations and could subsequently classify the last 28 AMM implementations without conflicts. 
The final AMM taxonomy consists of 16 AMM dimensions and 43 AMM characteristics.

To improve the usableness of the AMM taxonomy, we categorized its 16 dimensions into \textit{liquidity}, \textit{pricing}, and \textit{trading}. We inductively developed these groups from the dimensions included in the AMM taxonomy.

After the fourth iteration, we met the ending conditions (see Table~\ref{tab:methods-ending-cond}). Thus, we decided that the AMM taxonomy is final.

\subsection{AMM Archetype Development}

Building on our AMM taxonomy, we developed AMM archetypes by identifying dimensions that systematically differentiate AMM designs.
We compared the AMMs classified in the taxonomy to identify initial differentiating dimensions.
All authors independently rated the level to which each dimension helps discriminate between AMM designs.
For example, `token price source' received high ratings as a distinguishing factor because some AMMs determine prices internally (price-discovering), while others adopt external prices (price-adopting).

Next, we discussed the individual ratings among the authors to reach a unanimous consensus on the level to which the dimension is useful to discriminate between AMM designs.
The comparison revealed four dimensions as discriminatory: `source of liquidity', `token price source', `price discovery mechanism', and `liquidity concentration'.
For other dimensions, such as `parameter adjustment', the comparison did not uncover systematic differences between AMM designs. Another example is the dimension `risk management', which did not differentiate the analyzed AMM implementations.

Based on the four differentiating dimensions and their subordinate characteristics, we grouped the AMMs analyzed in taxonomy development (see Section~\ref{sec:amm-taxonomy-method}).
We identified 22 preliminary AMM archetypes based on combinations of the selected AMM characteristics.

We recognized that many characteristics used to develop the preliminary AMM archetypes use similar approaches for sourcing liquidity and discovering token prices.
We selected `token price source' and `source of liquidity' as key differentiating dimensions.
By constructing the cross-product of the dimensions `source of liquidity' and `token price source', we developed four preliminary AMM archetypes: `Price-discovering LP-based AMM', `Price-discovering Supply-sovereign AMM', `Price-adopting LP-based AMM', and `Price-adopting Supply-sovereign AMM'.
Among the 110 analyzed AMMs, we found no real-world instances of the `Price-adopting Supply-sovereign AMMs.' Consequently, we removed it from the final archetype set.
The final three AMM archetypes mainly differ in their approaches to source liquidity and discover adequate token prices.

%% file: 01_content/23_taxonomy.tex
\section{Taxonomy of Automated Market Makers in Cryptoeconomic Systems}
\label{sec:amm-taxonomy} \label{sec:groups-dim-chars}
 
This section presents the AMM taxonomy along its three groups of dimensions (see~Table~\ref{tab:results-groups-amm-char}): liquidity, pricing, and trading. These groups include 16 dimensions with 43 subordinate characteristics of AMM designs. 
In addition, Supplementary Material~\ref{app:tamm-full} offers an overview of the taxonomy in tabular form. Supplementary Material~\ref{app:demo} demonstrates how AMMs can be classified into the AMM taxonomy.

\begin{table}[b]
    \centering
    \caption{Overview of groups of AMM characteristics}
    \label{tab:results-groups-amm-char}
    \begin{tabularx}{\textwidth}{|l|X|}
    \hline
       \textbf{Group} & \textbf{Name}  \\
       \hline
       Liquidity & The source, availability, and constraints of liquidity \\
       \hline
       Pricing & The algorithms and properties important to set token exchange rates \\
       \hline
        Trading & The types, functionalities, and properties of trade execution  \\    
       \hline   
    \end{tabularx}
\end{table}

\subsection{Liquidity}
The \textit{liquidity} group focuses on token availability in liquidity pools to enable seamless trading.
AMM designs directly influence the available liquidity through liquidity pool structure, token pair configurations, and incentive models for LPs.

\subsubsection{Number of Tokens per Liquidity Pool}
\textit{The variety of tokens that can be deposited in one liquidity pool.}
The number of exchangeable token types (e.g., Bitcoin, Cardano, and Ether) per liquidity pool can be exactly two or more.

\italicparagraph{Influence}
The number of tokens per liquidity pool influences the flexibility of token exchange. For example, in a liquidity pool with two tokens, market participants can exchange a token \textit{A} for a token \textit{B} and a token \textit{B} for a token \textit{A}. In a liquidity pool with three token types, market participants can exchange in six directions: $A \leftrightarrow B$, $A \leftrightarrow C$, and $B \leftrightarrow C$.
More available exchange routes increase the variety of possible transactions, which can increase the capital efficiency of LPs' liquidity positions.

\subsubsection{Risk Management}
\textit{The mechanisms to manage the risk of holding volatile tokens.}

AMMs are exposed to the risk of inventory imbalance (see Section \ref{sec:amm-foundation}). AMMs can have no risk management, imbalance surcharges, or loss insurance. AMMs with no risk management entirely pass the risk to LPs. AMMs with imbalance surcharges add surcharges to stated bid/ask token prices based on the current inventory imbalance in order to reduce imbalance over time.
Imbalance surcharges can, however, disincentivize market participants to execute transactions, increasing the imbalance. AMMs that implement loss insurance compensate for losses of LPs with external tokens (e.g., the AMM's own tokens like UNI or SUSHI).

\italicparagraph{Influence}
Effective risk management enhances the appeal of AMMs for LPs, leading to increased liquidity and higher trading volumes. Strategies such as imbalance surcharges or loss insurance incentivize LPs by reducing risks and improving the stability and efficiency of the AMM. By addressing inventory imbalances and mitigating the effects of volatile token prices, risk management strategies can enhance market quality and support AMMs in maintaining liquidity during price fluctuations~\cite{kirste_influence_2024}.

\subsubsection{Source of Liquidity}
\textit{The origin of tokens that AMMs utilize for executing trades with market participants and supplying liquidity to markets.}

The source of liquidity for an AMM can be external LPs or internal token supply sovereignty.
In external liquidity provision, external LPs deposit tokens into liquidity pools of the AMM. In return, LPs are commonly rewarded with a share of the trading fee.

In internal liquidity provision, AMMs have supply sovereignty over tokens and do not depend on external LPs.
AMMs with internal liquidity provisions mint new tokens when market participants buy; AMMs burn tokens when market participants sell the tokens to the AMM.

\italicparagraph{Influence}
An AMM's liquidity source directly affects its ability to provide liquidity and maintain operational efficiency. AMMs relying on external LPs must incentivize them to deposit tokens, making them vulnerable to liquidity shortfalls during periods of volatility or declining rewards. This reliance necessitates carefully calibrated incentive mechanisms to retain liquidity while also increasing exposure to systemic risks like LP-based concentrated liquidity or sudden withdrawals~\cite{kirste_influence_2024, cartea_decentralised_2022}.
In contrast, an internal source of liquidity through token supply sovereignty solves the LAP by design. By ensuring continuous liquidity without reliance on external LPs, supply sovereignty enables full control over liquidity levels and guarantees a lower bound for liquidity~\cite{kirste_undergirding_2025, rouviere_continuous_2017}.

\subsubsection{Supported Token Pairs} 
\textit{The token pairs that can be traded against each other using an AMM.} 

AMMs can be open to all tokens and allow arbitrary token trading pairs. AMMs can have restricted token pairs that require tokens to be paired with another (e.g., USDC paired with any other token) or restrict the tokens that can be paired (e.g., USDC, BTC, and WETH).

\italicparagraph{Influence}
The token pairs supported by an AMM influence its flexibility. Restriction of token pairs increases the demand for individual tokens. For example, suppose a cryptoeconomic system has its own token. All token pairs in the AMM must be paired with the proprietary token. This increases the demand for the proprietary token because when LPs provide liquidity, they must buy the proprietary token as well. Limiting the supported token pairs to specific tokens, however, leads to more permissioned AMMs.

\subsection{Pricing}

The group \textit{pricing} focuses on the algorithms, mechanisms, and properties that are crucial to compute and adjust token exchange rates. 

\subsubsection{Information Incorporation} 
\textit{The approach for incorporating market information into prices.}

Price discovery mechanisms can be incorporative or non-incorporative. Price discovery mechanisms are incorporative when they extract information from transactions of market participants with the AMM. When market participants buy tokens, the token price is assumed to be too low. The AMM increases the token price to approximate an adequate token price~\cite{abernethy_optimization-based_2011, engel_composing_2021, oneill_can_2021}.
Non-incorporative price discovery mechanisms do not adjust token prices based on the transactions of market participants but receive prices from external sources like price oracles.

\italicparagraph{Influence}
Information incorporation into price discovery mechanisms strongly influences the ability of AMMs to perform accurate and efficient price discovery.
Incorporative AMMs use the market participants' transactional behavior directly in the price adjustment process, making token prices responsive to real-time market conditions. This dynamic adjustment ensures that token prices approximate adequate token prices by reflecting aggregated market information, such as perceived value and demand shifts. Consequently, information-incorporative AMMs operate independently of external price sources.

Non-incorporative AMMs typically use external token price sources, making AMMs dependent on the accuracy and reliability of token prices stated by those sources.
Such dependence can make AMMs vulnerable—for example, they may be susceptible to manipulated price feeds, delays in price updates, and inaccuracies during periods of high volatility.
Non-incorporative AMMs may suffer from market inefficiencies, as token prices fail to reflect real-time supply and demand dynamics driven by market participants' transactions.

\subsubsection{Liquidity Concentration}
\textit{The distribution of liquidity across token prices.}

AMMs can use function-based liquidity concentration, LP-based liquidity concentration, or autonomous liquidity concentration~\cite{cartea_decentralised_2022, xu_sok_2023, khakhar_delta_2022}. 
AMMs with function-based liquidity concentration use mathematical functions that prescribe the available liquidity at each token price. AMMs with LP-based liquidity concentration allow LPs to define their own price ranges to concentrate their liquidity~\cite{hayden_uniswap_2021, cartea_decentralised_2022}.
AMMs with autonomous liquidity concentration automatically concentrate the liquidity in token price ranges based on dynamic mathematical functions~\cite{hulsmans_dynamic_2021}. Such AMMs often use time-weighted average prices to determine the token price ranges where liquidity is concentrated~\cite{wang_automated_2020, cartea_decentralised_2022}. 

\italicparagraph{Influence}
Liquidity concentration in AMMs influences the available liquidity at different token prices, which in turn affects token price stability and evolution.
For example, suppose that liquidity for a correlated token pair (e.g.,~USDC/USDT) is concentrated at a token price of 1~USDC per 1~USDT. The AMM can settle large transaction volumes with small token price changes. If the token price increases to 10~USDC per 1~USDT, liquidity decreases, and token prices become more sensitive to changes.
LP-based liquidity concentration allows LPs to directly affect token price development by adjusting how sensitive token prices are within specific price ranges~\cite{drossos_automated_2025}.
Autonomous liquidity concentration implements proprietary mechanisms that determine at which token prices liquidity is concentrated.

\subsubsection{Liquidity Sensitivity}
\textit{The extent to which liquidity in an AMM affects the magnitude of token price changes.}

Liquidity-sensitive AMMs couple token price changes with available liquidity. 
In contrast, liquidity-insensitive AMMs decouple token price changes from available liquidity.

\italicparagraph{Influence}
Liquidity sensitivity in AMMs directly affects token price adjustments. In liquidity-sensitive AMMs, high liquidity dampens price fluctuations and reduces slippage, making them well-suited for large trades in dynamic markets. Conversely, low liquidity amplifies price changes, increasing slippage. In liquidity-insensitive AMMs, token price changes are independent of available liquidity \cite{othman_practical_2013, wang_automated_2020, cartea_decentralised_2022}. While this decoupling can simplify pricing mechanisms, it may lead to inefficiencies in price discovery and token valuation.

\subsubsection{Path Deficiency}
\textit{The feasible transactions in relation to the token reserve of an AMM.}

Path deficiency guarantees that token reserves are bounded from below (e.g.,~token reserves cannot shrink) for any set of transactions. Optimal transactions change the AMM token reserves to the minimal reserve set that satisfies its token reserve lower bound. Transactions above the lower bound either receive less output or add more input as the optimal transaction.

Strict path deficiency leads all transactions to be sub-optimal for the market participant. Market participants receive less output or add more input than optimal transactions. This ensures that the AMM's token reserves increase through trades.
Path-independent AMMs are path deficient by definition~\cite{angeris_improved_2020}.
Strictly path-deficient AMMs are path-dependent~\cite{angeris_improved_2020} because market participants must overpay the AMM to increase its token reserves. The AMM transitions to different states when a buy transaction and a sell transaction with equal volume are executed.

\italicparagraph{Influence}
Path deficiency influences resilience of AMMs, long-term sustainability, and incentives for market participants.
By ensuring that token reserves are always bounded from below, path deficiency prevents reserve depletion, allowing the AMM to maintain its functionality even during periods with high trading volume~\cite{angeris_improved_2020}.

Strict path deficiency forces market participants to execute sub-optimal trades.
This increases token reserves over time, bolstering the AMM's ability to withstand price shocks. However, this mechanism can increase trading costs for market participants, making path-deficient AMMs less attractive to them. If fewer market participants use the AMM, fewer trades are executed. Thus, the AMM's liquidity decreases.

\subsubsection{Path Independence}
\textit{The independence of AMMs' state transitions from the order of buy and sell transactions with identical cumulative volume.}

Pricing mechanisms can be path-dependent or path-independent.
Rearranging the order of buy and sell transactions with identical cumulative volumes leads to path-dependent pricing mechanisms to transition to different AMM states.
Path-dependent pricing mechanisms can state different token prices for the same transaction depending on its position in the transaction sequence~\cite{othman_practical_2013, wang_automated_2020, bartoletti_theory_2022}.
For example, buying 100~USDT in a single transaction increases the token price of the trading pair by 1; executing ten transactions with each 10 USDT volume increases the token price of the pair only by $0.5$. 
Path-independent pricing mechanisms transition to the same state for different sequential orders of transactions with identical cumulative volumes. Following the previous example, buying 100 USDT---irrespective of executing single or multiple transactions---always results in a token price increment of 1.

\italicparagraph{Influence}
Path independence can influence transaction volumes and trading behavior~\cite{abernethy_optimization-based_2011}.
Path-dependent mechanisms call for splitting transactions into smaller volumes to decrease costs, particularly when fees are non-linear with respect to volume. This behavior results in a trade-off between reducing transaction costs and incurring higher execution costs from additional operations~\cite{angeris_improved_2020}.

For path-independent AMMs, market participants often execute transactions in full to minimize infrastructure costs, such as gas fees.
Path independence also influences market efficiency by reducing the strategic complexity of transaction sequencing. Additionally, path-independent AMMs update state updates with less computational complexity, improving scalability~\cite{angeris_improved_2020}.

\subsubsection{Price Discovery}
\textit{The algorithms used for discovering adequate token prices.}

AMMs use eight principal price discovery algorithms: constant product algorithm, geometric mean algorithm, constant sum algorithm, constant product-sum algorithm, constant power sum algorithm, logarithmic market scoring algorithm, exponential function, and price adoption algorithm.
Each algorithm relies on a distinct conservation function, a mathematical equation that maintains an invariant relationship between token reserves in a liquidity pool. These functions govern how token prices adjust in response to supply and demand dynamics within the AMM.

A \textit{constant product} algorithm uses a conservation function based on a constant product~\cite{angeris_improved_2020, zinsmeister_uniswap_2020, xu_sok_2023}:

\begin{eqnarray}
c = \Pi_{i=1}^{n}r_i
\end{eqnarray}

$r$ indicates the amount of token $i$ in reserve. $n$ refers to the number of different token types tradable against the AMM.
Constant product algorithms facilitate dynamic price discovery across a wide range, theoretically spanning $[0, \infty]$. However, in practice, liquidity constraints influence the attainable price range.

\vspace{0.25cm}
\textit{Geometric mean} algorithms add the capability to change balances of token reserves~\cite{evans_liquidity_2021, zanger_g3msgeneralized_2022}. The conservation function of the geometric mean algorithm is defined as:

\begin{eqnarray}
 c = \Pi_{i=1}^{n}r_{i}^{w_i}   
\end{eqnarray}

$w$ indicates to the weight of the token reserves.
Geometric mean algorithms allow the creation of liquidity pools that are imbalanced by design, allowing LPs to have a higher exposure toward one token. Furthermore, geometric mean algorithms typically support more than two tokens in liquidity pools. This allows LPs to diversify their liquidity positions to earn more liquidity rewards.

\vspace{0.25cm}
\textit{Constant sum} algorithms enforce constant exchange rates~\cite{zanger_g3msgeneralized_2022, xu_sok_2023, mohan_automated_2022, angeris_improved_2020}. They use conservation functions that are based on constant sums:

\begin{eqnarray}
    c = \sum_{i=1}^{n} r_i
\end{eqnarray}

Constant sum algorithms influence other AMM characteristics.
For example, constant sum algorithms typically have fixed token exchange rates, making them impervious to trade volume, information-non-incorporative, and liquidity-insensitive. This constrains their use case to the fixed-ratio exchange of highly correlated tokens.
Furthermore, it should be noted that these algorithms can be completely depleted of one token, rendering them unable to facilitate exchanges in that token.

\vspace{0.25cm}
\textit{Constant product-sum} algorithms use a conservation function based on a constant sum of products. The conservation function can be expressed as:

\begin{eqnarray}
 \chi D^{x_i} + \prod x_i = \chi D^n + (\frac{D}{n})^n   
\end{eqnarray}

$D$ indicates the total amount of coins. $\chi$ indicates a leverage factor, which is defined as:
${\chi \in \mathbb{R} | 0 \leq \chi}$.

For $\chi=0$, the constant product-sum algorithm approximates a constant product conservation function. 
or $\chi=\infty$, the constant-sum algorithm approximates a constant sum conservation function\cite{egorov_stableswap_2019, egorov_automatic_2021}.

Constant product-sum algorithms combine the characteristics of constant-product and constant-sum algorithms.
Constant product-sum algorithms can be configured to exhibit minimal price changes around a specified exchange ratio, similar to constant-sum algorithms, while still being able to discover any token price, as with constant-product algorithms. When token prices deviate from the specified ratio, however, price changes increase.

\vspace{0.25cm}
\textit{Constant power-sum} algorithms use a conservation function based on a constant sum of multiple powers. The conservation function can be expressed as:

\begin{eqnarray}
 c=\sum_{i=1}^{n} r_i^{1-t}   
\end{eqnarray}

$t$ is a parameter to change the curvature of the conservation function~\cite{niemerg_yieldspace_2020}. Constant power-sum algorithms use constant-sum algorithms, enabling discovery of any token price while having minimal price changes at a specified exchange ratio. They are particularly well-suited for highly correlated tokens, allowing fine-grained control over slippage and creating a low-volatility exchange environment that can be tuned to a preferred rate.

\vspace{0.25cm}
\textit{Logarithmic market scoring} algorithms use a cost function $C$ of the total of assets in the market that can be expressed as follows: 
\begin{eqnarray}
 C(q) = b\cdot log(\sum_{j=1}^{n}exp(q_j/b))   
\end{eqnarray}
with $q$ being the vector of quantities, $b$ being a strictly positive parameter to control liquidity in the market and $n$ the number of assets~\cite{hanson_combinatorial_2003, slamka_prediction_2013, carvalho_logarithmic_2022, carvalho_permissioned_2020}. Logarithmic market scoring algorithms are particularly well-suited for prediction markets by bounding potential loss and managing risk for prediction tokens. This enables continuous liquidity and real-time price discovery even with limited participation.

\vspace{0.25cm}
\textit{Price adoption} algorithms use external price oracles to adopt token prices, which are then adjusted by a mathematical function~\cite{dodoex_dodo_2023}. Typically, price adoption algorithms adjust token prices based on their token reserve imbalances. Token reserve imbalances can be expressed as follows:

\begin{eqnarray}
 \Delta R_i = \frac{r_{0,\text{target}}-r_{0,\text{current}}}{r_{0,\text{target}}}   
\end{eqnarray}

$r_{0,\text{target}}$ indicates the targeted token reserves of $r_0$.  $r_{0,\text{current}}$ represents the actual token reserves of $r_0$.
The offered token price can be expressed as follows: 

\begin{eqnarray}
P_i=p_{\text{adopted}}+p_{\text{adopted}}k(\Delta R_i)
\end{eqnarray}

$k$ is a parameter to configure the magnitude of token price adjustments, which is defined as {$k \in \mathbb{R} | 0 \leq k \leq 1$} \cite{dodoex_dodo_2023, xu_sok_2023}.
Price adoption algorithms rely on external price oracles to obtain adequate token prices.
By adopting externally sourced prices, these algorithms reduce reliance on arbitrage traders to correct mispricings. This mechanism can lower slippage and improve price efficiency, though it introduces risks tied to oracle accuracy and manipulation, which can lead to financial losses for LPs.
Furthermore, smaller or emerging token markets may lack suitable token price oracles since most price feeds originate from large exchanges (e.g., Binance, Coinbase, or Uniswap), which typically list only well-established tokens.

\vspace{0.25cm}

\textit{Exponential function} pricing algorithms use an exponential conservation function based on a constant exponent~\cite{zargham_curved_2020, blockscience_alpha_2022, lau_single_2018}:

\begin{eqnarray}
 c= S^\kappa_a / r_b   
\end{eqnarray}

$S_a$ indicates the total token supply $a$. $r_b$ denotes the amount of token $a$ in reserve. $\kappa$ is a curvature parameter that influences price sensitivity to changes in reserve balances. A higher $\kappa$ leads to steeper price adjustments, making liquidity more responsive to demand shifts.
Exponential function pricing algorithms are typically tailored to specific use cases.

\subsubsection{Token Price Source}
\textit{The source that an AMM uses to quote bid/ask adequate token prices.}

The token price source can be internal or external. AMMs with internal token price sources use a built-in price discovery component to compute appropriate token prices~\cite{angeris_improved_2020,mohan_automated_2022}.
AMMs with external price sources outsource the discovery of adequate token prices to price oracles~\cite{dodoex_dodo_2023}.

\italicparagraph{Influence}
The token price source strongly affects the efficiency and responsiveness of AMMs in maintaining adequate token prices. External price sources, such as oracles, enable AMMs to adjust token prices without requiring direct buy/sell transactions. Reliance on external sources, however, introduces dependencies on the used oracle~\cite{pfister_finding_2022}, for example, in terms of the accuracy of stated token prices.
Such dependencies can facilitate price manipulation and delays in price updates.

In contrast, AMMs with internal token price sources discover adequate token prices using trade indications, such as buy and sell transactions~\cite{angeris_improved_2020}.
When new market information (e.g.,~fundamentals, regulatory news, new products) is available~\cite{fama_efficient_1970}, the token price stated by the AMM can diverge from the adequate token price. This divergence incentivizes rational market participants to collect and analyze market information to buy temporarily underpriced tokens and sell them in the future to realize profits~\cite{torres_frontrunner_2021, angeris_improved_2020}. Such market participants are called arbitrageurs. By continuously adopting token prices, they increase market efficiency. However, they can also erode the overall value in liquidity pools, potentially leading to financial losses for LPs.

\subsubsection{Translation Invariance}
\textit{The payoff from a portfolio consisting of equal amounts of each asset.}

% WELCHE CHARACTERISTICS GIBT ES
AMMs can be translation-invariant or translation-variant.
Translation-invariant AMMs always charge the same cost for the same amount of each asset \cite{schlegel_axioms_2022, othman_practical_2013, carvalho_logarithmic_2022, agrawal_unified_2009}. For example, a translation-invariant AMM charges 0.5~USD for token $A$ and 0.5~USD for token $B$. The cost of one token $A$ and one token $B$ (same amount) is 1~USD. Later, the AMM's token prices change to 0.8~USD for token $A$ and 0.2~USD for token $B$. The cost of one token $A$ and one token $B$ is 1~USD. The translation-invariant AMM always charges 1~USD for one token $A$ and one token $B$ at any state of the AMM.

Translation-variant AMMs charge different costs for the same amount of each asset at different states of the AMM~\cite{schlegel_axioms_2022}. A non-translation-invariant AMM charges, for example, 0.5~USD for token $A$ and 0.5~USD for token $B$. The cost of one token $A$ and one token $B$ (same amount) is 1~USD. Later, the AMM's token prices change to 0.8~USD for token $A$ and 0.3~USD for token $B$. The cost of one token $A$ and one token $B$ is 1.10~USD. Translation-variant AMMs charge different costs for one token $A$ and one token $B$ at different states of the AMM. 

\italicparagraph{Influence}
Translation invariance influences predictability and stability of pricing structures in AMMs. Translation invariant AMMs provide consistent pricing for liquidity positions with equal amounts of tokens, regardless of state changes~\cite{schlegel_axioms_2022}. This predictability simplifies portfolio management and hedging strategies for market participants, as the cost of balancing liquidity positions is constant over time~\cite{milionis_automated_2022}.

In contrast, translation variance leads to variability in the cost of a balanced liquidity position, which can create uncertainty for market participants. Such variability may incentivize strategic trading to exploit price differences but can also lead to inefficient management of liquidity positions~\cite{schlegel_axioms_2022}.

\subsubsection{Volume Dependence} 
\textit{The dependency of token prices on transaction volumes.}

Pricing mechanisms of AMMs can be volume-dependent or volume-independent. For volume-dependent pricing mechanisms, transaction volume leads to different mean token prices per transaction.
Volume-independent pricing mechanisms state equal mean token prices independent of transaction volumes. 

\italicparagraph{Influence}
Volume dependence directly impacts market efficiency, trading behavior, and liquidity management in AMMs. In volume-dependent pricing mechanisms, traders may be incentivized to split large trades into smaller ones to minimize slippage and reduce costs~\cite{othman_practical_2013}.
Splitting large-volume transactions can increase the number of processed transactions within a short time, leading to higher processing costs (e.g., gas fees).

\subsection{Trading}
The group \textit{trading} refers to the types, functionalities, and properties of trade execution via an AMM.

\subsubsection{Interoperability} \textit{The capability of an AMM to execute transactions across DLT systems.}

AMMs can be interoperable or non-interoperable.
Interoperable AMMs enable market participants to transact across DLT systems~\cite{sunyaev_token_2021}. Non-interoperable AMMs only settle transactions in cryptoeconomic systems built on a single DLT system. Senders and recipients of tokens must use the same DLT systems.

% Influence
\italicparagraph{Influence}
Interoperability enhances the accessibility and flexibility of AMMs by allowing market participants to transact seamlessly across multiple DLT systems. This capability reduces friction associated with transacting assets between DLT systems, enabling more efficient trade execution and more efficient liquidity management. Interoperable AMMs also attract a larger user base by bridging fragmented markets and facilitating cross-system transactions \cite{pfister_finding_2022}, enhancing overall market efficiency.

Non-interoperable AMMs offer simplicity, lower operational complexity, and reduced security risks. By operating within a single DLT system, non-interoperable AMMs avoid coordination and technical overhead required for cross-system transactions. This can result in faster transaction processing and lower transaction costs~\cite{kannengieser_bridges_2020}.

\subsubsection{Limit Order Functionality} \textit{The functionality to create limit orders.}

AMMs can either support limit order functionality or not.
Limit orders are instructions to buy and sell tokens at specified prices~\cite{abergel_limit_2016}. AMMs that support limit orders maintain order books where such instructions are recorded. AMMs execute limit orders when the token price reaches the specified limit.

\italicparagraph{Influence}
Limit order functionality of AMMs enables market participants to define price levels at which trades are executed, offering greater control over transaction outcomes. This functionality is particularly valuable for market participants who require precision in managing price exposure and mitigating market impact.
Limit order books can enhance liquidity of AMMs by efficiently matching opposing buy and sell orders at specified price levels.
However, offering limit order functionality increases the complexity of AMM design and infrastructure, potentially leading to higher operational costs and longer transaction settlement times.

\subsubsection{Parameter Adjustment}
\textit{The functionality to change AMM parameters after deployment.}

AMMs can have constant parameters (e.g., token weights), allow for manual parameter adjustments, or allow for automated parameter adjustments.
Constant parameters cannot be changed after deployment of the AMM~\cite{xu_sok_2023, churiwala_qlammp_2022}.
Manual adjustment allows specified market participants to change parameters. AMMs with automated adjustment use mechanisms that automatically adjust parameters. One exemplary parameter that is automatically adjusted is `trading fee'---a fee that AMMs charge for trade processing~\cite{xu_sok_2023, fan_towards_2022}

\italicparagraph{Influence}
AMM parameters can influence price discovery and available liquidity. For example, token weights influence amounts of individual tokens held in liquidity pools of AMMs. This can influence the available liquidity for the individual tokens.
Adjustment of trading fees can influence the bid/ask spreads. Such adjustments could also be used to mitigate potential attacks on AMMs. For example, front-running attacks during high token price volatility can be mitigated by increasing trading fees when token price volatility increases~\cite{xu_sok_2023}.

% \subsubsection{Price Guarantee}
% \textit{The process that determines token prices in cryptoeconomic system markets.}

% Token prices may change in the time between issuing transactions to a cryptoeconomic system and the settlement of those transactions. Market participants may thus sell or buy tokens at a different price than actually intended.
% AMMs can be designed to deal with changing token prices in three ways: give no price guarantee, guarantee price ranges, or guarantee exact prices. First, AMMs without a token price guarantee do not guarantee any price for transaction settlement. The settlement price is unknown at transaction issuance. Second, AMMs can implement ranged price determination that guarantees transaction settlement only in a specified price range, known as slippage. The slippage of token prices can usually be set per transaction. Third, AMMs can guarantee exact price determination. The token price at transaction issuance equals the token price at transaction settlement.

% % Influence
% The price guarantee of an AMM influences the ease of transaction execution for market participants. Configurable price ranges and exact price guarantees reduce the risk of transaction execution for market participants. Little price guarantees increase financial risks for market participants because transactions could be settled at unprofitable prices. 

%% file: 01_content/24_archetypes.tex
\newpage
\section{Archetypes of Automated Market Makers}
\label{sec:amm-archetypes}

We developed three AMM archetypes: Price-discovering LP-based AMM, Price-adopting LP-based AMM, and Price-discovering Supply-sovereign AMM.
The archetypes are mainly distinguished by two dimensions of the AMM taxonomy (i.e., \textit{price discovery} and \textit{source of liquidity}).
The following describes the three AMM archetypes.

%%%%%%%%%%%%%%%%%%%%%%%%%
%%%%%%%%%%%%%%%%%%%%%%%%%
\subsection{Price-discovering LP-based AMM}

\begin{figure}[h]
    \centering
    \includegraphics[width=0.7\linewidth]{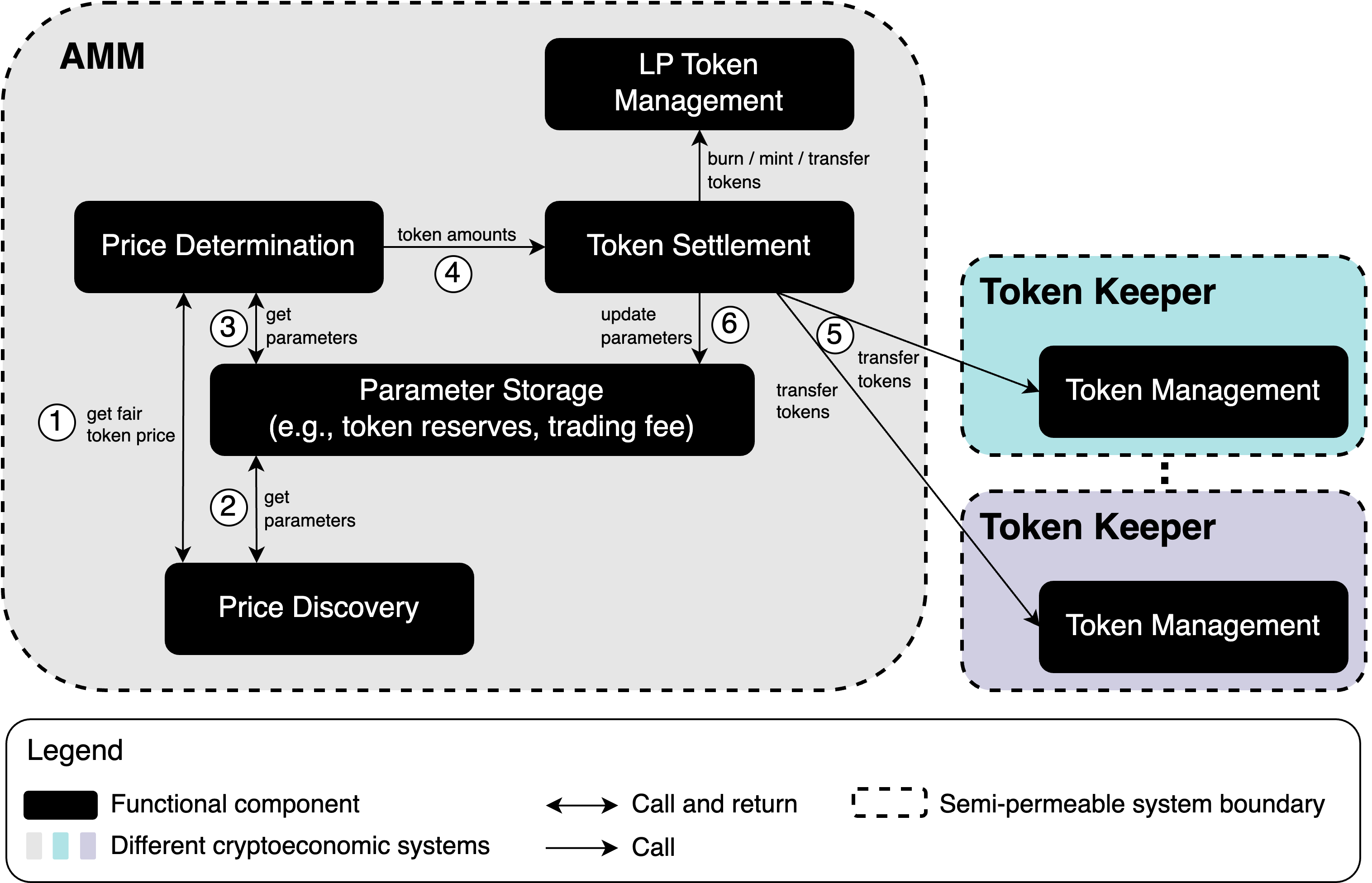}
    \caption{Architecture of the Price-discovering LP-based AMM and process of trade execution}
    \label{fig:amm-overview-pd-lp-amm}
\end{figure}

The Price-discovering LP-based AMM (see Figure~\ref{fig:amm-overview-pd-lp-amm}) has five internal key components: a price discovery component, a price determination component, a token settlement component, a parameter component, and an LP token management component.
The Price-discovering LP-based AMM trades tokens of token management components that at least one external token keeper operates.
The Price-discovering LP-based AMM is characterized by an \textit{internal price discovery} and \textit{LP-based source of liquidity}. 

The price discovery component of the Price-discovering LP-based AMM is initialized with a token price set by the first LP that initializes the AMM. Then, the price discovery component computes token prices based on a mathematical function that incorporates market participant perceptions of the token price as follows. Market participants only buy undervalued tokens and sell overvalued tokens to take profits.
When a market participant buys/sells tokens, it is assumed that the market participant knows information by which the new adequate token price diverges from the stated token price.
When market participants buy tokens, the Price-discovering LP-based AMM increases the token price because it is assumed that market participants perceive the adequate token price as higher than the token price stated by the AMM. Conversely, selling suggests a lower perceived fair price, leading to downward adjustments~\cite{angeris_improved_2020, evans_liquidity_2021}.
The magnitude of price changes depends on the liquidity available to the Price-discovering LP-based AMM. High liquidity commonly entails small token price changes, while low liquidity entails large token price changes..

The price discovery mechanism of the Price-discovering LP-based AMM is typically deterministic. Thus, price changes resulting from transactions of a given volume are predictable based on the AMM's state.
The price determination component of the Price-discovering LP-based AMM calls the price discovery component to fetch the current token price (steps~1 and 2), adjusts it by adding trading fees defined in the parameter component (step~3), and passes the determined token amounts to the token settlement component (step~4).

The token settlement component calls external token management components to transfer tokens between the Price-discovering LP-based AMM and market participants (step~5) and updates the parameters in the parameter component (step~6). Token management components record the token balances of the AMM and other market participants. The balances of the Price-discovering LP-based AMM in the token management components form the token reserves of its liquidity pool. The token management components are controlled by token keepers that can be part of multiple cryptoeconomic systems at the same time. Thus, tokens managed by token keepers could be accessed by multiple AMMs and other market participants.

LPs deposit tokens into liquidity pools in the token management components of external token keepers.
LPs receive LP tokens from the LP token keeper component in return for their token deposits. LP tokens represent a claim on a share of tokens in the Price-discovering LP-based AMM's liquidity pool. 
Using the LP tokens, LPs can withdraw their share of tokens from liquidity pools. By depositing and withdrawing tokens, the liquidity provided by the Price-discovering LP-based AMM can change over time.

The Price-discovering LP-based AMM uses an incentive mechanism to accumulate sufficient liquidity and keep LPs motivated not to withdraw deposited tokens. Common incentive mechanisms distribute market making revenues for token deposits to LPs. Such revenues correspond to shares of the transaction fees charged by the AMM for transaction settlement.

\paragraph{Use Cases}
The Price-discovering LP-based AMM is primarily used for\textit{decentralized token exchanges} and \textit{token issuances}.
The choice of price discovery mechanisms and liquidity sensitivity in this archetype is tied to the token types traded.
For decentralized token exchanges, the Price-discovering LP-based AMM can be used in correlated token markets, uncorrelated token markets, non-fungible token markets, and prediction markets.

\textit{Uncorrelated Token Markets:} For markets with uncorrelated tokens, Price-discovering LP-based AMMs typically use constant-product price discovery mechanisms. This allows the price discovery component of the AMM to set any token price. The optional LP-based liquidity concentration can increase the available liquidity at LP-chosen price ranges~\cite{hayden_uniswap_2021}.
Exemplary Price-discovering LP-based AMMs in uncorrelated token markets are Uniswap~v2~\cite{zinsmeister_uniswap_2020}, Uniswap~v3~\cite{hayden_uniswap_2021}, PancakeSwap~\cite{pancakeswap_docs_pancakeswap_2023}, and SushiSwap~\cite{sushiswap_sushi_2023}.

\textit{Correlated Token Markets:} For markets with correlated tokens, Price-discovering LP-based AMMs often use constant product-sum mechanisms~\cite{egorov_stableswap_2019, egorov_automatic_2021}. Combined with function-based liquidity concentration, that mechanism ensures high liquidity in tight price ranges, which is critical to ensure the exchange of correlated tokens~\cite{uniswap_uniswap_2023, sushi_docs_sushi_2023}. 
Exemplary Price-discovering LP-based AMMs in correlated token markets are Curve Finance~\cite{egorov_stableswap_2019} and mStable~\cite{mstable_pools_2023}.

\textit{Non-fungible Token Markets:} For markets with non-fungible tokens, Price-discovering LP-based AMMs typically use constant-product price discovery mechanisms. Due to limited NFT supply, these AMMs use discrete price steps instead of continuous pricing.
Exemplary Price-discovering LP-based AMMs in non-fungible token markets are Sudoswap~\cite{sudoswap_what_2024} and Hadeswap~\cite{hadeswap_liquidity_2022}.

\textit{Prediction Markets:}
Since prediction markets represent probabilities, the price must remain between 0 and 1, with price movements reflecting changes in perceived likelihood. This reflects the probability of an event occurring, with the AMM's price discovery mechanism adjusting these probabilities.
Exemplary Price-discovering LP-based AMMs in prediction markets are Augur~\cite{peterson_augur_2018}, Zeitgeist~\cite{zeitgeist_docs_prediction_2022}.

For token issuance, Price-discovering LP-based AMMs support initial token offerings. Price-discovering LP-based AMMs facilitate distribution of new tokens in markets. The token issuer, often developers of a cryptoeconomic system, initially provides liquidity by depositing the new tokens into the AMM's liquidity pool. The AMM then facilitates the exchange of these tokens, helping to establish their market value, and they enter circulation.
Exemplary Price-discovering LP-based AMMs for initial token offerings are Uniswap~v2~\cite{zinsmeister_uniswap_2020}, Uniswap~v3~\cite{hayden_uniswap_2021}, and Balancer~\cite{balancer_liquidity_2024}.

%%%%%%%%%%%%%%%%%%%%%%%%%
%%%%%%%%%%%%%%%%%%%%%%%%%
\subsection{Price-adopting LP-based AMM}
\begin{figure}[h]
    \centering
    \includegraphics[width=0.7\linewidth]{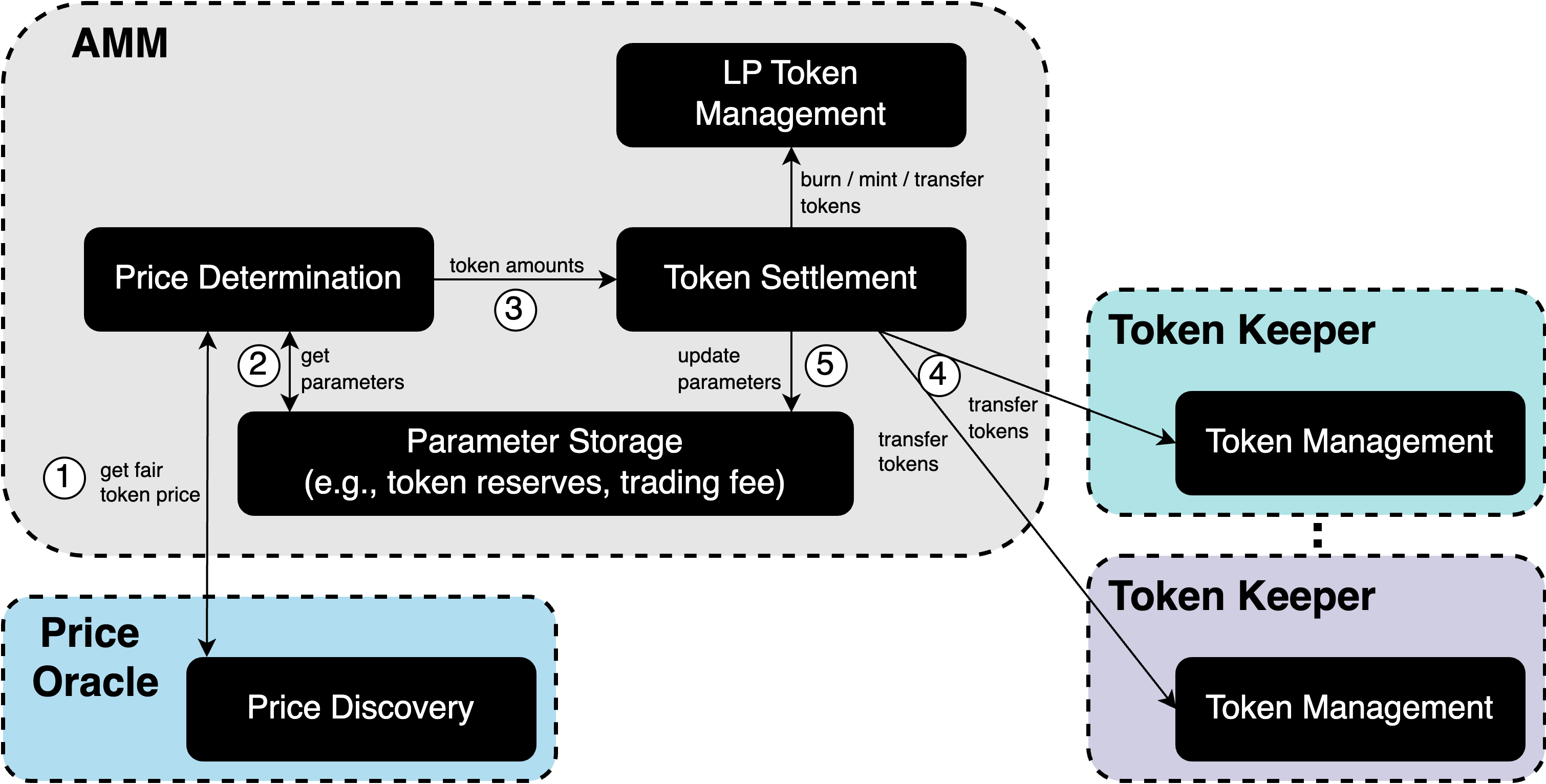}
    \caption{Architecture of the Price-adopting LP-based AMM and process of trade execution}
    \label{fig:amm-overview-pa-lp-amm}
\end{figure}

The Price-adopting LP-based AMM (see Figure \ref{fig:amm-overview-pa-lp-amm}) has four internal key components: a price determination component, a token settlement component, a parameter component, and an LP token management component.
The Price-adopting LP-based AMM uses at least one external price discovery component, which is part of the token keeper of other cryptoeconomic systems. The traded tokens are managed by external token management components. 
The Price-adopting LP-based AMM is characterized by an \textit{external price discovery} and \textit{LP-based source of liquidity}. 

\newpage
The price determination component requests token prices from at least one external price discovery component (step~1) that is part of an external \textit{price oracle} (e.g.,~Chainlink). The price oracle discovers adequate token prices in a proprietary way (e.g.,~retrieving token prices from Binance or Coinbase). Received token prices can be modified by the price determination component based on parameters of the parameter component (step~2). For example, the Price-adopting LP-based AMM commonly adds a surcharge to the token price when its token reserves become imbalanced. Given equal transaction volumes, the token reserves become less imbalanced in percentage terms when having high liquidity compared to low liquidity. Thus, in the case of high liquidity, the surcharge remains smaller. In consequence, high liquidity decreases the token price sensitivity. Low liquidity increases the token price sensitivity. 
The adjusted token price is subsequently passed from the price determination component to the token settlement component (step~3).

The Price-adopting LP-based AMM's token settlement component interacts with external token management components that manage tokens and record balances of market participants (step~4). The token settlement component then updates the parameters in the parameter component (step~5). The LP token management component issues LP tokens to the LPs, presenting a claim on a share of tokens in the liquidity pool of the Price-adopting LP-based AMM. Through depositing and withdrawing token deposits, the AMM's available liquidity can change over time.
The Price-adopting LP-based AMM implements an incentive mechanism that distributes revenues for token deposits to incentivize market participants to deposit tokens into the liquidity pools of the Price-adopting LP-based AMM.

\paragraph{Use Cases}

The Price-adopting LP-based AMM is used for \textit{decentralized token exchange} in correlated token markets, uncorrelated token markets, non-fungible token markets, and prediction markets~\cite{dodoex_dodo_2023, woofi_woofi_2023}.
While the fundamental design remains similar across these use cases, the parameters in the parameter component vary depending on the price correlation of traded tokens. For example, correlated tokens are likely to have smaller surcharges since token reserves are less prone to divergence, reducing the risk that LPs accumulate less valuable tokens. In contrast, uncorrelated tokens tend to have higher surcharges due to price fluctuations, increasing potential risks for LPs. because token prices fluctuate, increasing potential risks for LPs.

Examples of Price-adopting LP-based AMMs are DODO~\cite{dodoex_dodo_2023} and WOOFi~\cite{woofi_woofi_2023}.

%%%%%%%%%%%%%%%%%%%%%%%%%
%%%%%%%%%%%%%%%%%%%%%%%%%
\subsection{Price-discovering Supply-sovereign AMM}
\label{sec:price-disc-sup-sov}

\begin{figure}[h]
    \centering
    \includegraphics[width=0.7\linewidth]{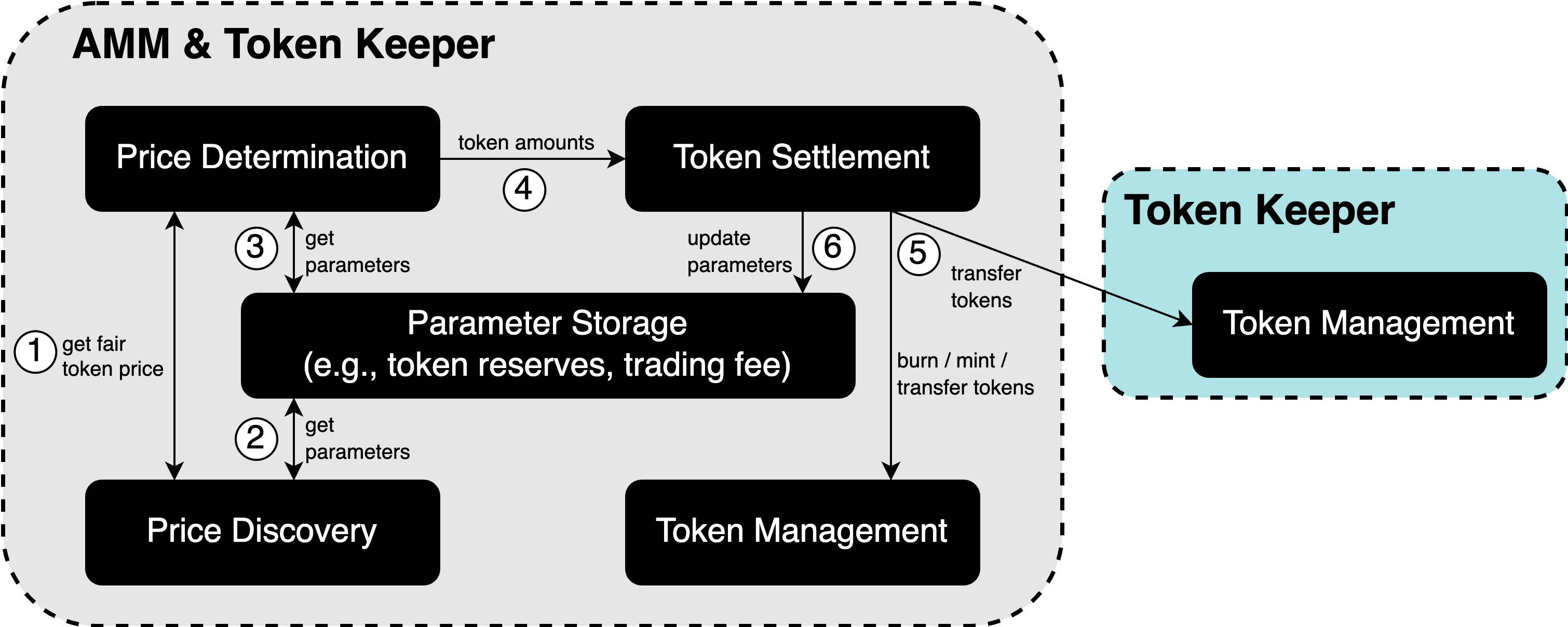}
    \caption{Architecture of the Price-discovering Supply-sovereign AMM and process of trade execution}
    \label{fig:amm-overview-pd-ss-amm}
\end{figure}

The Price-discovering Supply-sovereign AMM (see Figure~\ref{fig:amm-overview-pd-ss-amm}) has five internal key components: a price discovery component, a price determination component, a parameter component, a token settlement component, and a token management component. 
The Price-discovering Supply-sovereign AMM incorporates at least one token management component, allowing it to transfer, mint, and burn its own tokens because the AMM functions as its own token keeper. In contrast to the previous two AMM archetypes, the Price-discovering Supply-sovereign AMM does not incorporate an LP token management component.
The Price-discovering Supply-sovereign AMM is characterized by \textit{internal price discovery} and \textit{supply-sovereign source of liquidity}.

% internal price discovery
When the Price-discovering Supply-sovereign AMM is designed, its owner (e.g.,~a software developer) defines a supply curve for the issued token within the price discovery component. The supply curve is a mathematical function describing the relationship between circulating token supply to a token price. Although arbitrary supply curve shapes are possible, it is common practice to use monotonically increasing supply curves. As for the Price-discovering LP-based AMM, market participant perceptions of the token price are incorporated. Market participants can buy/sell undervalued/overvalued tokens to profit from price divergences. The price discovery component uses the prescribed supply curve to change token prices~\cite{angeris_improved_2020}.
When market participants buy/sell tokens, the circulating token supply increases/decreases because tokens are minted/burned by the internal token management component.

The Price-discovering Supply-sovereign AMM regulates the circulating token supply by instructing its token management component to mint/burn tokens, which increases/decreases the amount of circulating tokens.
This mechanism enables a primary market for those tokens. Unlike the previous AMM archetypes, the Price-discovering Supply-sovereign AMM eliminates reliance on external LPs by establishing a reversible bond-to-mint mechanism that starts with zero tokens minted. When users buy tokens from the Price-discovering Supply-sovereign AMM, the paid tokens are bonded in the liquidity pool and available to be paid out when circulating tokens are sold back to the AMM. Thereby, liquidity is, by design, guaranteed to be available for the complete circulating supply due to the reversible bond-to-mint mechanism~\cite{zargham_curved_2020, bcrg_exploring_2023, rouviere_tokens_2017}. 
The bond-to-mint mechanism uses the defined supply curve for price discovery and is typically deterministic. Thus, price changes of transactions with a given volume are predictable. The price discovery component of the Price-discovering Supply-sovereign AMM passes the adequate token price to the price determination component (steps~1 and 2). The price determination component adjusts the token price based on parameters (e.g.,~trading fees) of the parameter component (step~3). The adjusted token price is subsequently passed to the token settlement component (step~4).

% supply-sovereign source of liquidity
The token settlement component calls the internal and external token management components (step~5) and updates the parameters in the parameter component (step~6). The internal token management component is called to burn/mint tokens. The external token management components transfer tokens between the Price-discovering Supply-sovereign AMM and market participants. 
In the following, we describe how the Price-adopting LP-based AMM can provide sufficient liquidity and discover adequate token prices.

\paragraph{Use Cases}

The Price-discovering Supply-sovereign AMM is mainly used for \textit{token issuance} because supply-sovereignty is a fundamental requirement.
Common purposes of token issuance are initial token offerings~\cite{hertzog_bancor_2018} and curation tokens~\cite{ocean_protocol_foundation_ocean_2020, the_graph_graph_2023}. For both purposes, issuers must define the supply curves implemented in the Price-discovering Supply-sovereign AMM.

Examples of Price-discovering Supply-sovereign AMMs include The Graph Curation~\cite{the_graph_graph_2023}, Ocean Protocol Curation~\cite{ocean_protocol_foundation_ocean_2020}, and the concept of Undergirding Bonding Curves which are especially suitable for continuous token offerings~\cite{kirste_undergirding_2025}.

% ADD UBC HERE

%\newpage

%% file: 01_content/25_disclusion.tex
\section{Discussion}
\label{lbl:discussion}

This work presents an AMM taxonomy, including three groups (i.e.,~pricing, liquidity, and trading) comprising 16 dimensions with a total of 43 characteristics (see Table \ref{tab:results-groups-amm-char}). 
Building on the AMM taxonomy, we developed three AMM archetypes (i.e., the Price-discovering LP-based AMM, the Price-adopting LP-based AMM, and the Price-discovering Supply-sovereign AMM) with different approaches for supplying liquidity and discovering adequate token prices, making them suitable for different use cases in cryptoeconomic systems.
In the following, we discuss our principal findings and explain this work's contributions to research and practice.
Moreover, we elucidate the limitations of this work and outline future research directions.

\subsection{Principal Findings}

The results of this study suggest that proper AMM design is especially crucial for ensuring sufficient liquidity and discovering adequate token prices. Inappropriate liquidity provisioning and price discovery can lead to the thin market problem~\cite{anderson_new_2007}, which exacerbates financial risks, for example, due to increased price volatility. 
While the AMM archetypes suit different use cases, each uses distinct approaches to overcoming the thin market problem.
This indicates that the context given by use cases must be carefully considered to tackle the thin market problem in cryptoeconomic system markets.

We recognized the dominance of gray literature, technical documentation, and blog articles on AMMs. AMM development is largely practice-driven, with innovation occurring through real-world implementations rather than theoretical research. This suggests a gap between industry advancements and academic analysis.
Boundaries between AMMs and other parts of cryptoeconomic systems are often not clearly defined. For example, Uniswap is described as a peer-to-peer protocol for decentralized token exchange that uses an AMM~\cite{uniswap_uniswap_2023}. In contrast, the official Uniswap v3 whitepaper says that Uniswap is a non-custodial AMM~\cite{hayden_uniswap_2021}. Such inconsistent wording makes it difficult to draw clear system boundaries for AMMs and their uses in cryptoeconomic systems. We argue that AMMs are market makers implemented as software agents that trade tokens with market participants at self-determined prices in an automated manner. The purposes of AMMs include decentralized token exchanges and token issuance. A decentralized exchange (DEX) is a decentralized marketplace that is the equivalent of traditional exchanges. DEXs allow market participants to trade tokens with each other.
AMMs are virtual market participants within DEXs that continuously offer token exchanges at stated bid/ask prices. DEXs can use different types of AMMs.
AMMs typically manage a single liquidity pool for token exchanges.

The classification of AMM into the AMM taxonomy shows substantial differences between used price discovery mechanisms.
One reason for using different price discovery mechanisms is their strong influence on token price evolution, and AMMs must use different price discovery mechanisms to suit their purposes (e.g., correlated or uncorrelated token markets).
For example, constant-product price discovery mechanisms are frequently used in exchanges of uncorrelated tokens because token prices are changed almost equally at all token prices~\cite{xu_sok_2023}. Constant-product-sum price discovery mechanisms are preferred for exchanges of correlated tokens because token prices are less sensitive at a specified exchange rate than constant-produce price discovery mechanisms~\cite{xu_sok_2023}. Such dynamics in the price evolution of different price discovery mechanisms on token price evolution are needed to enable AMMs to meet their individual purposes.

Many AMM designs are customized forks of code repositories of established AMMs, such as Uniswap~v2 and Uniswap~v3. Such forks of AMM code bases are partially modified (e.g., SushiSwap adding staking and liquidity mining to Uniswap v2 or PancakeSwap running on Binance Smart Chain instead of the Ethereum system). Such modifications lead to a large variance between AMMs classified into the AMM taxonomy (see Supplementary Material~\ref{app:demo}). 
For example, the dimensions \textit{allowed trading pairs}, \textit{trading fee adjustment}, and \textit{parameter adjustment} are only slightly modified in most forks of the AMM designs~(e.g., PancakeSwap and Sushiswap). The solution approaches for sourcing liquidity and discovering adequate token prices remain largely unchanged. Since AMM operators usually issue their own tokens (e.g., UNI, SUSHI), the operators can profit by selling these issued tokens. We suspect that this monetary interest has created a large number of new AMMs that have modified existing AMMs with little effort to create their own AMMs to profit from selling custom tokens. 

Despite the wide variability of characteristics that are easy to modify in AMM designs, the applicability demonstration of the AMM taxonomy (see Supplementary Material~\ref{app:demo}) shows that single characteristics per dimension have become dominant in AMM designs. For example, most AMM designs are non-translation invariant and information expressive based on liquidity~\cite{zinsmeister_uniswap_2020, sushiswap_sushi_2023, pancakeswap_docs_pancakeswap_2023}. Only a few AMM designs with constant-sum price discovery are translation invariant and information inexpressive~\cite{mstable_pools_2023, xu_sok_2023, mohan_automated_2022}. The dimensions \textit{information expressiveness}, \textit{translation invariance}, and \textit{volume dependency} appear to depend on the dimension \textit{price discovery}. For example, AMMs with constant-sum price discovery are \textit{information inexpressive}, \textit{translation invariant}, and \textit{volume independent}. Such AMMs have known weaknesses for the purpose of decentralized token exchange. For example, due to \textit{information in-expressiveness} and \textit{volume independence}, they cannot adjust token prices~\cite{schlegel_axioms_2022, mohan_automated_2022}. In addition, AMMs with constant-sum price discovery can be depleted, making them unable to trade. mStable is the only AMM that used constant-sum price discovery until 2021 when they switched to constant-product-sum because of the weaknesses of constant-sum price discovery~\cite{mohan_automated_2022}.

We identified the Price-discovering LP-based AMM as the predominant archetype that is widespread, with 98 occurrences in the analysis of 110 AMMs (see Supplementary Material~\ref{app:demo}). Extant literature also focuses on Price-discovering LP-based AMM. The Price-discovering Supply-sovereign AMM and Price-adopting LP-based AMM are much less represented in extant literature. 
The Price-discovering LP-based AMM appears frequently due to forks of well-known AMMs, such as Uniswap v2, Uniswap v3, and Curve, which are attributed to the Price-discovering LP-based AMM. Because those forks change a few characteristics with minor influences on their approaches for sourcing liquidity and discovering adequate token prices, there is a large variance in the analyzed AMM designs. 

The Price-adopting LP-based AMM is less often used, presumably because of its dependency on external price oracles and its insufficient capabilities for discovering adequate token prices in newly emerging markets. However, we uncovered that the Price-adopting LP-based AMM can enable a cost-efficient decentralized token exchange if used in thick markets with reliable external price oracles. AMM archetypes with internal price discovery mechanisms suffer financial losses if token prices diverge from adequate token prices because tokens are sold under price or bought over price. These financial losses can significantly impact liquidity, potentially reaching a threshold where liquidity provisioning becomes unprofitable~\cite{kirste_influence_2024, drossos_automated_2025}.

The Price-discovering Supply-sovereign AMM is exclusively used for token issuance because it requires supply sovereignty over at least one token. The results indicate that the Price-discovering Supply-sovereign AMM excels in token issuance, allowing issuers to define supply curves for predictable price evolution and improved financial risk assessment. However, the Price-discovering Supply-sovereign AMM has scarcely been researched. Specific effects on price evolution still remain unclear.  Price-discovering Supply-sovereign AMMs show great potential for improving token issuance and exchange in cryptoeconomic systems. Their mechanisms to guarantee value, prescribe available liquidity and establishing a lower price bound are highly desirable. They could form a novel type of market structure that enables developers to create more sustainable and self-sovereign cryptoeconomic systems~\cite{kirste_undergirding_2025}.

\subsection{Contributions to Practice and Research}

Developing AMMs that meet specific use case requirements is challenging, requiring a deep understanding of their design characteristics and their technical and economic impact on cryptoeconomic systems. This study provides a conceptual foundation---an AMM taxonomy and AMM archetypes---that enhances understanding of AMM designs, such as in terms of liquidity sourcing and token price discovery. These insights enable software engineers to better meet requirements from economics (e.g., sufficient liquidity provisioning) and software design (e.g., secure discovery of token prices).
In particular, we contribute to research and practice in three ways.
First, we present an AMM taxonomy that describes key dimensions and characteristics in AMM designs. The AMM taxonomy can guide AMM development by enabling comparisons between AMM design options to meet software requirements.
Through such comparisons, developers can make better-informed decisions on the design of AMMs.

Second, by clarifying main influences of different AMM characteristics on cryptoeconomic systems, we bridge the previously disconnected perspectives of software engineering and economics on AMMs. This can help developers estimate whether envisioned AMMs will better meet use case requirements when selecting one characteristic over another, supporting better-informed AMM design.

Third, we present three AMM archetypes: Price-discovering LP-based AMM, Price-adopting LP-based AMM, and Price-discovering Supply-sovereign AMM. These archetypes strongly differ in fundamental characteristics of AMM designs, making them suitable for specific use cases and unsuitable for others. 
By explaining how AMM archetypes suit principal use cases, the AMM archetypes serve as blueprints for custom AMM designs. 
Together, the AMM taxonomy and archetypes offer a conceptual foundation for designing AMMs that align with diverse use cases (e.g., decentralized token exchanges and token issuance), facilitating more efficient and sustainable markets in cryptoeconomic systems.

\subsection{Limitations and Future Research}

We analyzed AMMs based on related official whitepapers (e.g., \cite{hayden_uniswap_2021, simpson_fetch_2019, ocean_protocol_foundation_ocean_2020}), blog articles (e.g.,~\cite{balasanov_bonding_2020, lau_single_2018, rouviere_introducing_2019}), and official documentation (e.g.,~\cite{dodoex_dodo_2023, uniswap_uniswap_2023}). Most of the analyzed publications are not peer-reviewed. Moreover, many publications do not present sufficiently detailed information to classify AMMs into the AMM taxonomy. To still classify all AMMs into the taxonomy, we inferred missing details by analyzing source code and referencing related AMM designs where applicable. For example, if the AMM is on a fork of established AMMs, we used publications on the original AMM design to complete the classification. While we documented and justified these assumptions, we acknowledge that some classifications remain uncertain.

Extant research is mainly concerned with Price-discovering LP-based AMMs (e.g.,~~constant-function market makers~\cite{xu_sok_2023, angeris_analysis_2021, mohan_automated_2022}) and Price-adopting LP-based AMMs (e.g., proactive market makers~\cite{dodoex_dodo_2023, mohan_automated_2022, xu_anatomy_2019}). 
Research on Price-discovering Supply-sovereign AMMs remains in its early stages. Only a few publications on Price-discovering Supply-sovereign AMMs are available~\cite{kirste_undergirding_2025, zargham_curved_2020, rouviere_tokens_2017}. Influences on cryptoeconomic system markets have been investigated in recent studies~\cite{kirste_undergirding_2025} and uncover desirable characteristics of Price-discovering Supply-sovereign AMMs.
However, such AMMs are rarely used in practice, requiring further research and economic reasoning to better understand the implications of such novel market structures with unlimited supply.

The AMM taxonomy and archetypes show that AMMs mainly differ in the mechanisms used for liquidity provisioning and price discovery.
We extracted the solution approaches implemented in AMM archetypes for sourcing liquidity and discovering adequate token prices based on literature and AMMs used in cryptoeconomic systems. We outline how AMM archetypes facilitate liquidity provisioning and price discovery.
Recent quantitative analyses have measured the influence of some AMM archetypes on the available liquidity and compared it to conventional market makers~\cite{kirste_influence_2024}. Price-discovering LP-based AMM with LP-based liquidity concentration can provide liquidity levels comparable to conventional market makers on Binance and Coinbase. For other AMM archetypes, however, we could not find empirical evidence for the efficacy of the developed archetypes in providing sufficient liquidity and reliably discovering adequate token prices.
Quantitative analyses should be performed for all AMM archetypes to compare their efficacy in sourcing liquidity and discovering adequate token prices.

In terms of liquidity provisioning, LP-based AMMs strongly depend on LPs that must be incentivized to deposit sufficient tokens to maintain high liquidity. Commonly, AMMs incentivize LPs by offering liquidity rewards \cite{cartea_decentralised_2022, drossos_automated_2025}. Such incentive mechanisms, however, seem insufficient as LPs commonly lose value instead of gaining it \cite{cartea_decentralised_2022}.
Future studies should explore novel incentive mechanisms that ensure profitability for both AMMs and LPs while maintaining economic sustainability, such as by optimizing trading fees to attract LPs without discouraging traders through too high fees.

\section{Conclusion}
\label{lbl:conclusion}

AMMs have become fundamental for issuing and trading tokens in cryptoeconomic systems. However, proper design of AMMs for cryptoeconomic systems is difficult because AMMs must meet conventional software requirements and additional economic requirements related to provide sufficient liquidity and trade at adequate token prices. 
To support the development of AMMs that enable smooth trading, this work presents an AMM taxonomy and three AMM archetypes (i.e.,~Price-discovering LP-based AMM, Price-adopting LP-based AMM, and Price-discovering Supply-sovereign AMM) with different solution approaches to source liquidity and discover adequate token prices.

From our perspective, supply-sovereign AMMs align well with the core principle of cryptoeconomic systems to enable more decentralized markets where market participants have sovereignty over their owned assets. They can potentially mitigate price volatility and guide token price evolution, eventually creating novel market structures enabled through the use of smart contracts. We believe that supply-sovereign AMMs have the potential to greatly supporting the viable operation of cryptoeconomic systems.

\newpage

%% file: 01_content/26_appendix.tex
\newpage
\section*{Supplementary Material}
\setcounter{table}{0}
\renewcommand{\thetable}{A\arabic{table}}

\section{Overview of the Dimensions and Characteristics in the AMM Taxonomy}
\label{app:tamm-full}

Table~\ref{tab:groups-amm-char} offers an overview of the developed AMM taxonomy.

\input{01_content/33_taxonomy_table}

\newpage
\clearpage

\section{Applicability Demonstration of the AMM Taxonomy}
\label{app:demo}

Table~\ref{tab:tamm-demo-neu} demonstrates how AMMs can be classified into the AMM taxonomy.
Each AMM has exactly one characteristic of each dimension. The characteristics are mutually exclusive because all AMMs have exactly one characteristic per dimension. The relevance and representativeness are shown because all characteristics are used by at least one AMM. The limited number of dimensions and characteristics shows the conciseness and exhaustiveness of the AMM taxonomy.

\input{01_content/31_demonstration_table}

\newpage

%% file: 01_content/33_taxonomy_table.tex
\begin{table}[h]
\centering
\caption{Overview of the AMM taxonomy}
\label{tab:groups-amm-char}

\begin{tabular}{|c|l|p{7cm}|l|} 
\hline
\textbf{\rotatebox[origin=l]{90}{Group}} & \textbf{Dimension} & \textbf{Description} & \textbf{Characteristic} \\ 

\hline
\multirow{26}{*}{\textbf{\rotatebox[origin=l]{90}{Pricing}}} & \multirow{2}{*}{\begin{tabular}[c]{@{}l@{}}Information \\ Incorporation Pool\end{tabular}} & \multirow{2}{*}{\parbox{7cm}{The approach for incorporating market information into prices.}} & Incorporative \\ 
\cline{4-4}
 & & & non-incorporative \\ 
\cline{2-4}
 & \multirow{3}{*}{Liquidity Concentration} & \multirow{2}{*}{\parbox{7cm}{The distribution of liquidity across token prices.}} & Automatic \\ 
\cline{4-4}
 & & & Function-based \\ 
\cline{4-4}
 & & & LP-based \\ 
\cline{2-4}
 & \multirow{2}{*}{Liquidity Sensitivity} & \multirow{2}{*}{\parbox{7cm}{The extent to which liquidity in an AMM affects the magnitude of token price changes.}} & Insensitive \\ 
\cline{4-4}
 & & & Sensitive \\ 
\cline{2-4}
 & \multirow{2}{*}{Path Deficiency} & \multirow{2}{*}{\parbox{7cm}{The feasible transactions in relation to the token reserve of an AMM.}} & Deficient \\ 
\cline{4-4}
 & & & Strictly Deficient \\ 
\cline{2-4}
 & \multirow{2}{*}{Path Independence} & \multirow{2}{*}{\parbox{7cm}{The independence of AMMs' state transitions from the order of transactions with identical cumulative volume.}} & Path Dependent \\ 
\cline{4-4}
 & & & Path Independent \\ 
\cline{2-4}
%  & \multirow{3}{*}{Price Bounding} & \multirow{2}{*}{\parbox{7cm}{The functionality to limit token prices of the AMM.}} & Bounded from Above \\ 
% \cline{4-4}
%  & & & Bounded from Below \\ 
% \cline{4-4}
%  & & & Bounded from Above and Below \\ 
% \cline{2-4}
 & \multirow{8}{*}{Price Discovery} & \multirow{2}{*}{\parbox{7cm}{The algorithms used for discovering adequate token prices.}} & Constant-sum \\ 
\cline{4-4}
 & & & Constant-power-sum \\ 
\cline{4-4}
 & & & Constant-product \\ 
\cline{4-4}
 & & & Constant-product-sum \\ 
\cline{4-4}
 & & & Exponential Function \\ 
\cline{4-4}
 & & & Geometric Mean \\ 
\cline{4-4}
 & & & Logarithmic Market Scoring \\ 
\cline{4-4}
 & & & Price Adoption \\ 
\cline{2-4}
 & \multirow{2}{*}{Token Price Source} & \multirow{2}{*}{\parbox{7cm}{The source that an AMM uses to quote bid/ask adequate token prices.}} & External \\ 
\cline{4-4}
 & & & Internal \\ 
\cline{2-4}
 & \multirow{2}{*}{Translation invariance} & \multirow{2}{*}{\parbox{7cm}{The payoff from a portfolio consisting of equal amounts of each asset.}} & Non-translation Invariant \\ 
\cline{4-4}
 & & & Translation Invariant \\ 
 \cline{2-4}
 & \multirow{2}{*}{Volume Dependency} & \multirow{2}{*}{\parbox{7cm}{The dependency of token prices on transaction volumes.}} & Volume-dependent \\ 
\cline{4-4}
 & & & Volume-independent \\

\hline

\multirow{13}{*}{\textbf{\rotatebox[origin=l]{90}{Liquidity}}}  & \multirow{2}{*}{\begin{tabular}[c]{@{}l@{}}Number of Tokens per \\ Liquidity Pool\end{tabular}} & \multirow{2}{*}{\parbox{7cm}{The variety of tokens that can be deposited in one liquidity pool.}} & Two \\ 
\cline{4-4}
 & & & Three or More \\ 
\cline{2-4}
 & \multirow{3}{*}{Risk Management} & \multirow{2}{*}{\parbox{7cm}{The mechanisms to manage the risk of holding volatile tokens.}} & Imbalance Surcharges \\ 
\cline{4-4}
 & & & Loss Insurance \\ 
\cline{4-4}
 & & & No Risk Management \\ 
\cline{2-4}
 & \multirow{2}{*}{Source of Liquidity} & \multirow{2}{*}{\parbox{7cm}{The origin of tokens that AMMs utilize for executing trades with market participants.}} & External \\ 
\cline{4-4}
 & & & Internal \\ 
\cline{2-4}
 & \multirow{2}{*}{Supported Trading Pairs} & \multirow{2}{*}{\parbox{7cm}{The token pairs that can be traded against each other using an AMM.}} & Open \\ 
\cline{4-4}
 & & & Restricted \\ 

\hline

\multirow{8}{*}{\textbf{\rotatebox[origin=l]{90}{Trading}}} & \multirow{2}{*}{Interoperability} & \multirow{2}{*}{\parbox{7cm}{The capability of an AMM to execute transactions across DLT systems.}} & Interoperable \\ 
\cline{4-4}
 & & & Non-interoperable \\ 
\cline{2-4}
 & \multirow{2}{*}{Limit Order Functionality} & \multirow{2}{*}{\parbox{7cm}{The functionality to create limit orders.}} & Included \\ 
\cline{4-4}
 & & & Not Included \\ 
\cline{2-4}

\cline{2-4}

 & \multirow{3}{*}{Parameter Adjustment} & \multirow{2}{*}{\parbox{7cm}{The functionality to change AMM parameters after deployment.}} & Automatic \\ 
\cline{4-4}
 & & & Fixed \\ 
\cline{4-4}
 & & & Manual \\ 
\cline{2-4}

\hline

\end{tabular}

\end{table}

%% file: 01_content/31_demonstration_table.tex
\begin{table}[ht]
\centering
\caption{Demonstration of the AMM Taxonomy}
\label{tab:tamm-demo-neu}

\resizebox{\textwidth}{!}{%
\begin{tabular}{|c|l|l|c|c|c|c|c|c|c|c|c|c|c|c|c|} 
\hline
\textbf{\rotatebox[origin=l]{90}{Group}}  & \textbf{Dimension}  & \textbf{Characteristic}   & \rotatebox[origin=l]{90}{Augur\cite{peterson_augur_2018}} & \rotatebox[origin=l]{90}{Balancer WP\cite{martinelli_balancer_2019}} & \rotatebox[origin=l]{90}{Balancer MP\cite{balancer_managed_2023}} & \rotatebox[origin=l]{90}{Bancor ST\cite{hertzog_bancor_2018}} & \rotatebox[origin=l]{90}{Curve v1\cite{egorov_stableswap_2019}} & \rotatebox[origin=l]{90}{DODO\cite{dodoex_dodo_2023}} & \rotatebox[origin=l]{90}{mStable 2021\cite{mstable_pools_2023, mohan_automated_2022}} & \rotatebox[origin=l]{90}{Ref.Finance\cite{reffinance_reffinanace_2023}} & \rotatebox[origin=l]{90}{THORSwap Sb\cite{thorchain_liquidity_2023}} & \rotatebox[origin=l]{90}{Uniswap v2\cite{zinsmeister_uniswap_2020}} & \rotatebox[origin=l]{90}{Uniswap v3\cite{hayden_uniswap_2021}} & \rotatebox[origin=l]{90}{WooFi\cite{woofi_woofi_2023}} & \rotatebox[origin=l]{90}{YieldSpace\cite{niemerg_yieldspace_2020}}  \\ 
\hline
\multirow{26}{*}{\textbf{\rotatebox[origin=l]{90}{Pricing}}}   & \multirow{2}{*}{\begin{tabular}[c]{@{}l@{}}Information \\ Incorporation\end{tabular}}  & Incorporative  & x   & x   & x   & x & x  & x & & x   & x  & x  & x  & x   & x   \\ 
\cline{3-16}
&  & Non-incorporative   & &  &  &   &  &   & x & &  & & & &  \\ 
\cline{2-16}
& \multirow{3}{*}{Liquidity Concentration} & Automatic & &  &  &   &  & x & & &  & & & x   &  \\ 
\cline{3-16}
&  & Function-based & x   & x   & x   & x & x  &   & x & & x  & x & & & x   \\ 
\cline{3-16}
&  & LP-based  & &  &  &   &  &   & & x   &  & & x  & &  \\ 
\cline{2-16}
& \multirow{2}{*}{Liquidity Sensitivity} & Insensitive & &  &  &   &  &   & x & &  & & & &  \\ 
\cline{3-16}
&  & Sensitive & x   & x   & x   & x & x  & x & & x   & x  & x  & x  & x   & x   \\ 
\cline{2-16}
& \multirow{2}{*}{Path Deficiency}    & Deficient & x   &  &  & x &  &   & & &  & & & &  \\ 
\cline{3-16}
&  & Strictly Deficient  & & x   & x   &   & x  & x & x & x   & x  & x  & x  & x   & x   \\ 
\cline{2-16}
& \multirow{2}{*}{Path Independence}  & Path Dependent & &  &  &   &  & x & & & x  & & & &  \\ 
\cline{3-16}
&  & Path Independent & x   & x   & x   & x & x  &   & x & x   &  & x  & x  & x   & x   \\ 
\cline{2-16}
& \multirow{3}{*}{Price Bounding} & Bounded from Above  & &  &  &   &  &   & & &  & & & & x   \\ 
\cline{3-16}
&  & Bounded from Above and Below  & & x   & x   &   & x  & x & & x   & x  & x  & x  & x   &  \\ 
\cline{3-16}
&  & Bounded from Below & &  &  &   &  &   & x & &  & & & &  \\ 
\cline{2-16}
& \multirow{8}{*}{Price Discovery}    & Constant-sum   & &  &  &   &  &   & x & &  & & & &  \\ 
\cline{3-16}
&  & Constant-power-sum & &  &  &   &  &   & & x   & x  & & & &  \\ 
\cline{3-16}
&  & Constant-product  & &  &  &   & x  &   & & &  & & & &  \\ 
\cline{3-16}
&  & Constant-product-sum  & &  &  &   &  &   & & &  & & & & x   \\ 
\cline{3-16}
&  & Exponential Function  & &  &  & x &  &   & & &  & & & &  \\ 
\cline{3-16}
&  & Geometric Mean & & x   & x   &   &  &   & & &  & & & &  \\ 
\cline{3-16}
&  & Logarithmic Market  Scoring  & x   &  &  &   &  &   & & &  & & & &  \\ 
\cline{3-16}
&  & Price Adoption & &  &  &   &  & x & & &  & & & x   &  \\ 
\cline{2-16}
& \multirow{2}{*}{Token Price Source} & External  & x   & x   & x   & x & x  &   & x & x   & x  & x  & x  & & x   \\ 
\cline{3-16}
&  & Internal  & &  &  &   &  & x & & &  & & & x   &  \\ 
\cline{2-16}
& \multirow{2}{*}{Translation Invariance}  & Non-translation Invariant & & x   & x   & x & x  & x & & x   & x  & x  & x  & x   & x   \\ 
\cline{3-16}
&  & Translation Invariant & x   &  &  &   &  &   & x & &  & & & &  \\ 
\cline{3-16}
& \multirow{2}{*}{Volume Dependency}  & Volume-dependent & x   & x   & x   & x & x  & x & & x   & x  & x  & x  & x   & x   \\ 
\cline{3-16}
&  & Volume-independent  & &  &  &   &  &   & x & &  & & & &  \\

\hline

\multirow{13}{*}{\textbf{\rotatebox[origin=l]{90}{Liquidity}}} 

& \multirow{2}{*}{\begin{tabular}[c]{@{}l@{}}Number of Tokens per \\ Liquidity Pool\end{tabular}} & Three or More    & &  &  & x & x  & x & x & x   & x  & x  & x  & x   & x   \\ 
\cline{3-16}
&  & Two  & x   & x   & x   &   &  &   & & &  & & & &  \\ 
\cline{2-16}
& \multirow{3}{*}{Risk Management}    & Imbalance Surcharges  & &  &  &   &  & x & & &  & & & x   &  \\ 
\cline{3-16}
&  & Loss Insurance & &  &  &   &  &   & & & x  & & & &  \\ 
\cline{3-16}
&  & No Risk Management  & x   & x   & x   & x & x  &   & x & x   &  & x  & x  & & x   \\ 
\cline{2-16}
& \multirow{2}{*}{Source of Liquidity}   & External  & x   & x   & x   &   & x  & x & x & x   & x  & x  & x  & x   & x   \\ 
\cline{3-16}
&  & Internal  & &  &  & x &  &   & & &  & & & &  \\ 
\cline{2-16}
& \multirow{2}{*}{Supported Trading Pairs} & Open   & x   & x   &  &   & x  & x & & x   & x  & x  & x  & x   &  \\ 
\cline{3-16}
&  & Restricted  & &  & x   & x &  &   & x & &  & & & & x   \\ 
\hline

\multirow{8}{*}{\textbf{\rotatebox[origin=l]{90}{Trading}}} & \multirow{2}{*}{Interoperability}   & Interoperable  & &  &  &   &  &   & & x   &  & & & x   &  \\ 
\cline{3-16}
&  & Non-interoperable   & x   & x   & x   & x & x  & x & x & & x  & x  & x  & & x   \\ 
\cline{2-16}
& \multirow{2}{*}{Limit Order Functionality}  & Included  & &  &  &   &  &   & & x   &  & & & &  \\ 
\cline{3-16}
&  & Not Included   & x   & x   & x   & x & x  & x & x & & x  & x  & x  & x   & x   \\ 
\cline{2-16}

& \multirow{3}{*}{Parameter Adjustment}  & Automatic & &  &  &   &  & x & & & x  & & & & x   \\ 
\cline{3-16}
&  & Fixed  & x   & x   &  & x & x  &   & x & &  & x  & x  & &  \\ 
\cline{3-16}
&  & Manual & &  & x   &   &  &   & & x   &  & & & x   &  \\ 
\cline{2-16}
\hline

\end{tabular}}

\end{table}